\documentclass[12pt, a4paper]{article}
\usepackage{cite}
\usepackage{amsmath,amssymb,multirow}
\usepackage[usenames,dvipsnames]{color}
\usepackage{bm}

\usepackage{graphicx}
\usepackage[bookmarksopen,colorlinks=true,linkcolor=dark_green,citecolor=dark_red,urlcolor=dark_blue,linktocpage=false]{hyperref}

\usepackage{comment}
\usepackage{slashed}

\definecolor{dark_blue}{rgb}{0,0,0.6}
\definecolor{red}{rgb}{1,0,0}
\definecolor{dark_green}{rgb}{0,0.4,0}
\definecolor{dark_red}{rgb}{0.6,0,0}

\makeatletter
\newcommand\footnoteref[1]{\protected@xdef\@thefnmark{\ref{#1}}\@footnotemark}
\makeatother

\usepackage[height=22.5cm,width=16.5cm,centering]{geometry}

\newcommand{\dd}{\mathrm{d}}

\def\thefootnote{\fnsymbol{footnote}}

\renewcommand{\thefootnote}{\fnsymbol{footnote}}
\newcommand{\Gcenter}[2]{
  \dimen0=\ht\strutbox
  \advance\dimen0\dp\strutbox
  \multiply\dimen0 by#1
  \divide\dimen0 by2
  \advance\dimen0 by-.5\normalbaselineskip
  \raisebox{-\dimen0}[0pt][0pt]{#2}}
\begin{document}
%%%%%%%%%%%%%%%%%%%%%%%%%%%%%%%%%%%%%%%%%%%%%%%%%%

%%%%%%%%%%%%%%%%%%%%%%%%%%%%%%%%%%%%%%%%%%%%%%%%%%
\begin{titlepage}

\begin{center}

\hfill UT-16-28\\
\hfill CTPU-16-27\\
\hfill KEK-TH-1927\\
\hfill IPMU-16-0134\\

\vskip .5in

{\LARGE \bf 
Violent Preheating in \\[.5em] Inflation with Nonminimal Coupling
}

\vskip .55in

{\large Yohei Ema$^a$, Ryusuke Jinno$^{b,c}$, Kyohei Mukaida$^d$ and Kazunori Nakayama$^{a,d}$}

\vskip 0.25in

\begin{tabular}{ll}
$^{a}$&\!\!\!\!\!\! {\em Department of Physics, Faculty of Science, The University of Tokyo,}\\
&\!\!\!\!\! {\em Bunkyo-ku, Tokyo 133-0033, Japan}\\[.3em]
$^{b}$ &\!\!\!\!\!\! {\em Center for Theoretical Physics of the Universe, Institute for Basic Science (IBS),}\\
&\!\!\!\!\! {\em Daejeon 34051, Korea}\\[.3em]
$^{c}$ &\!\!\!\!\!\! {\em Theory Center, High Energy Accelerator Research Organization (KEK),}\\
&\!\!\!\!\! {\em Oho, Tsukuba, Ibaraki 305-0801, Japan}\\[.3em]
$^{d}$ &\!\!\!\!\!\! {\em Kavli IPMU (WPI), UTIAS, The University of Tokyo,}\\
&\!\!\!\!\! {\em Kashiwa, Chiba 277-8583, Japan}
\end{tabular}

\end{center}
\vskip .45in

\begin{abstract}
\noindent
We study particle production at the preheating era in inflation models with nonminimal coupling
$\xi \phi^2R$ and quartic potential $\lambda \phi^4/4$
for several cases: real scalar inflaton, complex scalar inflaton and Abelian Higgs inflaton.
We point out that the preheating proceeds much more violently than previously thought.
If the inflaton is a complex scalar, the phase degree of freedom is violently produced at the first stage of preheating.
If the inflaton is a Higgs field, the longitudinal gauge boson production is similarly violent.
This is caused by a spike-like feature in the time dependence of the inflaton field,
which may be understood as a consequence of the short time scale during which the effective potential or kinetic term changes suddenly.
The produced particles typically have very high momenta $k \lesssim \sqrt{\lambda}M_\text{P}$.
The production might be so strong that almost all the energy of the inflaton is carried away 
within one oscillation for $\xi^2\lambda \gtrsim {\mathcal O}(100)$.
This may partly change the conventional understandings of the (p)reheating after inflation
with the nonminimal coupling to gravity such as Higgs inflation.
We also discuss the possibility of unitarity violation at the preheating stage.

\end{abstract}

\end{titlepage}

\tableofcontents
\thispagestyle{empty}

\renewcommand{\thepage}{\arabic{page}}
\setcounter{page}{1}
\renewcommand{\thefootnote}{$\diamondsuit$\arabic{footnote}}
\setcounter{footnote}{0}
%%%%%%%%%%%%%%%%%%%%%%%%%%%%%%%%%%%%%%%%%%%%%%%%%%

\newpage
\setcounter{page}{1}

%%%%%%%%%%%%%%%%%%%%%%%%%%%%%%%%%%%%%%%%%%%%%%%%%%
\section{Introduction}
\label{sec_introduction}
\setcounter{equation}{0}
%%%%%%%%%%%%%%%%%%%%%%%%%%%%%%%%%%%%%%%%%%%%%%%%%%

After the results from the Planck satellite, a simple chaotic inflation model
with a power-law potential~\cite{Linde:1983gd} is excluded or disfavored~\cite{Ade:2015xua}.
Hence large field inflation models need to be modified so that 
the prediction of the scalar spectral index and tensor-to-scalar ratio falls into the 
observationally favored region.

One of the simple ideas is to add a nonminimal gravitational coupling of the inflaton to the Ricci scalar:
\begin{align}
S
&= \int \dd^4x \sqrt{-g_J}
\left[ 
\left( \frac{M_\text{P}^2}{2} + \frac{\xi}{2}\phi_J^2 \right)R_J
- \frac{1}{2}g_J^{\mu \nu}\partial_\mu \phi_J \partial_\nu \phi_J
- V_J(\phi_J) 
\right],
\label{eq:B_SJ}
\end{align}
where the potential is given by
\begin{align}
V_J(\phi_J)
&= \frac{\lambda}{4}\phi_J^4.
\label{eq:B_V}
\end{align}
Here $M_\text{P}$ is the reduced Planck mass, 
$g_{J \mu\nu}$ is the metric,
$\phi_J$ is the inflaton,
$R_J$ is the Ricci scalar,
and $\xi$ and $\lambda$ are model parameters.\footnote{
	We attach the subscript $J$ to quantities in the Jordan frame 
	to distinguish them from those in the Einstein frame in this paper.
}
An interesting feature of this model is that
the potential becomes effectively flat in the large field value region,
providing a good candidate for the inflaton potential~\cite{Futamase:1987ua,Fakir:1990eg}.
Throughout this paper, we focus on this class of models.
A specific example is 
the so-called Higgs inflation~\cite{CervantesCota:1995tz,Bezrukov:2007ep,Bezrukov:2009db}, 
where the standard model (SM) Higgs boson plays the role of the inflaton.\footnote{
	The applicability of Higgs inflation after the discovery of Higgs boson at LHC 
	is found in Refs.~\cite{Degrassi:2012ry,Bednyakov:2015sca,George:2015nza}.
} This model is consistent with the observations when
the parameter satisfies $\xi \sim 5\times 10^{4}\sqrt{\lambda}$~\cite{Ade:2015xua},
and we assume $\xi \gg 1$ throughout this paper.
The quartic coupling for the Higgs inflation case, $\lambda \sim 0.01$, satisfies this condition,\footnote{
	It has been pointed out that in Higgs inflation $\xi \gg 1$ generically generates
	large $R^2$ term in the action without some fine-tuning~\cite{Salvio:2015kka}.
} though we do not limit ourselves to the SM Higgs field as the inflaton 
but consider more general scalar fields in this paper.
For example, a gauge-singlet scalar dark matter~\cite{Lerner:2009xg,Lebedev:2011aq},
or U(1)$_{\rm B-L}$ Higgs~\cite{Okada:2011en} as an inflaton has been considered.

After inflation ends, the universe enters the (p)reheating phase in which the inflaton field is rapidly oscillating 
around its potential minimum.
It is known that the first stage of the reheating after inflation is often accompanied with
explosive particle production, 
if either the inflaton strongly couples to other fields and/or the inflaton oscillation amplitude is large enough,
and this is called preheating~\cite{Traschen:1990sw,Shtanov:1994ce,Kofman:1994rk,Kofman:1997yn}.
Actually there may be several possible large couplings in the present model:
the nonminimal coupling, inflaton self-coupling, gauge coupling etc.
The (p)reheating of the Higgs inflation was studied in Refs.~\cite{Bezrukov:2008ut,GarciaBellido:2008ab,Bezrukov:2014ipa,Repond:2016sol},
where it was pointed out that gauge bosons are efficiently produced at the preheating stage.

In this paper we revisit the preheating after inflation with the nonminimal coupling.
We first analyze the background dynamics of the inflaton carefully, and find that there are two mass scales 
(or inverse time scales) 
in the inflaton oscillation for $M_\text{P}/\xi \ll \Phi \ll M_\text{P}$,
where $\Phi$ is the inflaton oscillation amplitude in the Einstein frame
(see Sec.~\ref{sec:Weyl} for the definition of the Einstein frame). One is the usual inflaton oscillation scale,
\begin{align}
	m_\mathrm{osc} = \Delta t_\mathrm{osc}^{-1} \sim \frac{\sqrt{\lambda}M_\text{P}}{\xi},
\end{align}
and the other is a much shorter time scale which we call the ``spike'' scale,
\begin{align}
	m_\mathrm{sp} = \Delta t_\mathrm{sp}^{-1} \sim \sqrt{\lambda}\Phi.
\end{align}
This time scale $\Delta t_\mathrm{sp}$ corresponds to the time interval at which the inflaton passes through 
the region with $\lvert \phi_J \rvert \sim \lvert \phi \rvert \lesssim M_\text{P}/\xi$,
where $\phi$ is the inflaton in the Einstein frame.
The spike time scale appears in the dynamics of $\phi_J$ and $\phi$.
The dynamics of $\phi_J$ imprints this fast time scale in the following reason.
$\phi_J$ and the scalar component of the metric have kinetic mixing due to the nonminimal coupling.
For $\lvert \phi_J \rvert \gtrsim M_\text{P}/\xi$, this mixing term dominates over the original inflaton kinetic term 
$-g^{\mu\nu}_J \partial_\mu \phi_J\partial_\nu \phi_J/2$,
and the kinetic term of the inflaton effectively changes at around $\lvert \phi_J \rvert \sim M_\text{P}/\xi$.
Thus, the time scale $\Delta t_\mathrm{sp}$ is induced in the dynamics of $\phi_J$ as a change of the kinetic term.
The dynamics of $\phi$ also imprints this fast time scale 
because the shape of the inflaton potential changes at around $\lvert \phi \rvert \sim M_\text{P}/\xi$.
It appears as a peculiar behavior in, \textit{e.g.} the effective inflaton mass in the Jordan frame
and the conformal factor when the inflaton passes through near the origin, which we call ``spike''-like feature.
This peculiar feature has long been overlooked in the literature 
except for a few studies~\cite{Tsujikawa:1999me,DeCross:2015uza},\footnote{
 	We will clarify differences of our study from them
	in Secs.~\ref{sec:real} and~\ref{sec:global_U1}, respectively.
} and has recently been investigated in detail by one of the present authors~\cite{J:kurorekishi}.
In this paper, we point out that this feature causes 
much more violent particle production than previously thought.
In particular, if the inflaton is gauge-charged, the production of the longitudinal gauge boson 
is significantly enhanced compared with that of the transverse gauge boson.
The difference in the behavior of the longitudinal and transverse modes
has already been pointed out in Ref.~\cite{Lozanov:2016pac} 
in the context of preheating without the nonminimal coupling,
and we see that this difference becomes a significant one
if the inflaton has a large nonminimal coupling to gravity.
The energy transfer to the longitudinal mode is so violent that
almost all the energy density of the inflaton can be transferred to the longitudinal gauge bosons
within one oscillation after the end of inflation.

The organization of this paper is as follows.
In Sec.~\ref{sec_background}, we analyze the background dynamics in the Einstein frame.
In Sec.~\ref{sec_pp}, we discuss particle production by this oscillating background,
taking real, global U(1) charged, and gauge U(1) charged inflaton as examples.
Sec.~\ref{sec_conc} is devoted to conclusions and discussion.

%%%%%%%%%%%%%%%%%%%%%%%%%%%%%%%%%%%%%%%%%%%%%%%%%%
\section{Background dynamics}
\label{sec_background}
\setcounter{equation}{0}
%%%%%%%%%%%%%%%%%%%%%%%%%%%%%%%%%%%%%%%%%%%%%%%%%%

In this section, we discuss the background dynamics of inflation models
with nonminimal coupling between the inflaton and the Ricci scalar as Eq.~(\ref{eq:B_SJ}).
In particular, we concentrate on the inflaton oscillation regime after inflation.
We consider a real scalar inflaton in this section, 
but the background dynamics is the same as a complex scalar case 
that we will discuss in the next section.

The action~\eqref{eq:B_SJ} is often called a ``Jordan frame'' action.
By performing the Weyl transformation, we can remove the nonminimal coupling, and
such a frame is called the ``Einstein frame.''
It may be simpler to analyze the inflaton dynamics in the Einstein frame,
but still couplings of the inflaton to other particles show a peculiar feature even in that frame
as we will see in Sec.~\ref{sec:mapping}.
This is mainly because the theory is defined in the Jordan frame, not in the Einstein frame,
and hence the other particles couple to $\phi_J$, not $\phi$.\footnote{
	More specifically, other particles couple to $\phi_J$ with a constant coupling that does not depend on $\phi_J$,
	and regard it as the tree level action.
	Whether or not this treatment is natural from the viewpoint of quantum field theory may be an interesting issue.
}
In the following, we first perform the Weyl transformation to see the relation between
the Einstein and Jordan frames in Sec.~\ref{sec:Weyl}.
We derive the equations of motion in each frame in Sec.~\ref{sec:eom}.
Then, we discuss how the fast mass scale $m_\mathrm{sp}$ appears in the dynamics 
of the inflaton in Sec.~\ref{sec:mapping}.
The last subsection is the most important part of this paper.

%%%%%%%%%%%%%%%%%%%%%%%%%%%%%%%%%%%%%%%%%%%%%%%%%%
\subsection{Weyl transformation}
\label{sec:Weyl}
%%%%%%%%%%%%%%%%%%%%%%%%%%%%%%%%%%%%%%%%%%%%%%%%%%

First we transform the action~\eqref{eq:B_SJ}
to that in the Einstein frame.
We perform the Weyl transformation of the metric
\begin{align}
g_{\mu \nu}
&\equiv \Omega^2 g_{J\mu \nu},
\label{eq:B_conformal}
\end{align}
where the conformal factor $\Omega$ is defined as
\begin{align}
\Omega^2
&= 1 + \frac{\xi \phi_J^2}{M_\text{P}^2}.
\label{eq:B_Omega}
\end{align}
Under this transformation, the Ricci scalar behaves as
\begin{align}
R_J
= \Omega^2
\left(
R + 
6\Box \ln \Omega - 
6g^{\mu \nu} 
\partial_\mu \ln \Omega \;
\partial_\nu \ln \Omega
\right),
\label{eq:R_Weyl}
\end{align}
where $R$ is the Ricci scalar constructed from $g_{\mu \nu}$, 
and $\Box \equiv g^{\mu \nu} \nabla_\mu \nabla_\nu$ with $\nabla_\mu$ 
denoting the covariant derivative associated with $g_{\mu\nu}$.
In order to canonically normalize the kinetic term,
we redefine the inflaton as
\begin{align}
\frac{\dd\phi}{\dd\phi_J}
&= \frac{1}{\Omega^2}\sqrt{1 + \frac{\xi(1 + 6\xi) \phi_J^2}{M_\text{P}^2}}.
\label{eq:B_dphi_tEdphi}
\end{align}
Then, the action is given by
\begin{align}
S
&= \int \dd^4x \sqrt{-g}
\left[ \frac{M_\text{P}^2}{2}R 
- \frac{1}{2}g^{\mu \nu} \partial_\mu \phi \partial_\nu \phi
-  V(\phi) 
\right],
\label{eq:B_SEcanonical}
\end{align}
where the potential is redefined as
\begin{align}
V(\phi)
&\equiv \frac{1}{\Omega^4}V_J(\phi_J).
\label{eq:B_VE}
\end{align}
This is the action in the so-called Einstein frame.
We can explicitly see the relation between the inflatons in the Jordan and
Einstein frames by solving Eq.~\eqref{eq:B_dphi_tEdphi}:
\begin{align}
\frac{\phi}{M_\text{P}}
&= 
\sqrt{\frac{1 + 6\xi}{\xi}}\sinh^{-1}\left( \frac{\sqrt{\xi(1 + 6\xi)}\phi_J}{M_\text{P}} \right)
- \sqrt{6}\sinh^{-1}\left( \sqrt{\frac{6\xi^2\phi_J^2/M_\text{P}^2}{1 + \xi \phi_J^2/M_\text{P}^2}} \right).
\label{eq:mapping}
\end{align}
For $\xi \gg 1$, it reduces to
\begin{align}
\frac{\phi}{M_\text{P}}
&\simeq 
\begin{cases}
\displaystyle \frac{\phi_J}{M_\text{P}} 
\;\;\;\; 
&{\rm for}~~~|\phi| \ll \displaystyle\frac{M_\text{P}}{\xi},
\\[.8em]
\displaystyle \sqrt{\frac{3}{2}} \ln \left( 1 + \frac{\xi \phi_J^2}{M_\text{P}^2} \right)
\;\;\;\; 
&{\rm for}~~~|\phi| \gg \displaystyle\frac{M_\text{P}}{\xi}.
\end{cases}
\label{eq:B_phi_tEphi_approx}
\end{align}
The latter relation also implies
\begin{align}
	\frac{\phi}{M_\text{P}} \simeq \sqrt{\frac{3}{2}} \frac{\xi\phi_J^2}{M_\text{P}^2}~~~~~~{\rm for}~~~~~~\frac{M_\text{P}}{\xi}\ll |\phi|\ll M_\text{P}.
\end{align}
The potential becomes
\begin{align}
V(\phi)
&\simeq 
\begin{cases}
\displaystyle \frac{\lambda}{4}\phi^4
\;\;\;\; 
&{\rm for}~~~|\phi| \ll \displaystyle\frac{M_\text{P}}{\xi},
\\
\displaystyle 
\frac{\lambda M_\text{P}^4}{4\xi^2}
\left[1 - \exp\left(-\sqrt{\frac{2}{3}}\frac{\phi}{M_\text{P}}\right) \right]^2
\;\;\;\; 
&{\rm for}~~~|\phi| \gg \displaystyle\frac{M_\text{P}}{\xi}.
\end{cases}
\label{eq:B_VE_approx}
\end{align}
In the intermediate region, the potential is well approximated by the quadratic one:
\begin{align}
	V(\phi) \simeq \frac{\lambda M_\text{P}^2}{6 \xi^2}\phi^2~~~~~~{\rm for}~~~~~~\frac{M_\text{P}}{\xi}\ll |\phi|\ll M_\text{P}.
\end{align}
Note that the potential is extremely flat in the large field value region with
$\lvert \phi \rvert \gg M_\text{P}$ where inflation takes place.

%%%%%%%%%%%%%%%%%%%%%%%%%%%%%%%%%%%%%%%%%%%%%%%%%%
\subsection{Equations of motion}
\label{sec:eom}
%%%%%%%%%%%%%%%%%%%%%%%%%%%%%%%%%%%%%%%%%%%%%%%%%%

Now we derive the background equations of motion 
in the Einstein and Jordan frames, respectively.

%%%%%%%%%%%%%%%%%%%%%%%%%%%%%%%%%%%%%%%%%%%%%%%%%%
\subsubsection*{Einstein frame}
%%%%%%%%%%%%%%%%%%%%%%%%%%%%%%%%%%%%%%%%%%%%%%%%%%

First we derive the background equations of motion in the Einstein frame.
We assume that the background metric is given by
the Friedmann-Lema\^itre-Robertson-Walker (FLRW) one:
\begin{align}
	\dd s^2
	&= - N^2(t) \dd t^2 + a^2(t) \dd x^i \dd x^i,
\end{align}
where $N$ and $a$ are the lapse function and the scale factor in the Einstein frame, respectively.
$i = 1,2,3$ is a spatial index and its summation is promised.
We also assume that the background part of the inflaton depends only on time, or $\phi = \phi(t)$.
By taking variation with respect to $\phi$, $N$ and $a$, we obtain the following background
equations of motion in the Einstein frame:\footnote{
\label{fn_redundant}
	One of them is redundant.
}
\begin{align}
	\ddot{\phi}
	+ 3H\dot{\phi} + \frac{\dd V}{\dd \phi}
	&=
	0,
	\label{eq:B_phi_tEOM} \\
	3M_\text{P}^2H^2
	&= \frac{1}{2} \dot{\phi}^2 + V,
	\label{eq:B_Friedmann_E} \\
	M_\text{P}^2\left( 2\dot{H} + 3H^2 \right)
	&= 
	- \frac{1}{2} \dot{\phi}^2 + V,
	\label{eq:B_Raychaudhuri_E}
\end{align}
where the dot denotes the derivative with respect to $t_E$
defined as $\dd t_E = N \dd t$,\footnote{
Although the lapse function $N$ is often set to be unity in the literature, we do not set $N = 1$ in this paper.
This is because 
one can clearly see that it is also Weyl-transformed,
which makes the relation between the Jordan frame 
and the Einstein frame manifest.
}
and $H = \dot{a}/a$ is the Hubble parameter.
Inflation occurs for $\lvert \phi \rvert \gg M_\text{P}$ in this model.
Once the field value of the inflaton becomes $\lvert \phi \rvert \lesssim M_\text{P}$,
it starts to oscillate around the origin of its potential.
The potential is quadratic and the mass scale of the oscillation 
is $m_{\rm osc}^2 \sim \lambda M_\text{P}^2/\xi^2$ for $\Phi \gtrsim M_\text{P}/\xi$,
where $\Phi$ is the inflaton oscillation amplitude.
We mainly study the inflaton oscillation regime with $M_\text{P}/\xi \lesssim \Phi \lesssim M_\text{P}$ in this paper.
The inflaton energy density in this regime is given by
\begin{align}
	\rho_\phi
	&\simeq \frac{\lambda M_\text{P}^2}{6\xi^2}\Phi^2.
\end{align}
In the numerical estimation in the next section, we mainly adopt 
the model parameter $\lambda = 0.01$ and $\xi = 10^4$ 
and the initial condition $\phi = M_\text{P}$ with vanishing velocity.
With these parameter values, we find $\rho_\phi \simeq C_\phi \lambda M_\text{P}^4/\xi^2$
with $C_\phi \simeq 2\times10^{-2}$ around the first zero-crossing of the inflaton.
This numerical factor does not significantly change for other parameters.

%%%%%%%%%%%%%%%%%%%%%%%%%%%%%%%%%%%%%%%%%%%%%%%%%%
\subsubsection*{Jordan frame}
%%%%%%%%%%%%%%%%%%%%%%%%%%%%%%%%%%%%%%%%%%%%%%%%%%

Next we derive the equations of motion in the Jordan frame.
We take the inflaton to be $\phi_J = \phi_J(t)$, and
also the metric as the FLRW one:
\begin{align}
\dd s^2
&= - N_J^2(t)\dd t^2 + a_J^2(t)\dd x^i\dd x^i.
\end{align}
They are related to the metric in the Einstein frame as $N_J = N_E/\Omega$ and $a_J = a_E/\Omega$.
By taking a variation with respect to $\phi_J$, $N_J$ and $a_J$, we obtain the background equations of motion as\footnoteref{fn_redundant}
\begin{align}
	\frac{\dd^2\phi_J}{\dd t_J^2} + 3H_J \frac{\dd\phi_J}{\dd t_J} - \xi 
	\left( 6\frac{\dd H_J}{\dd t_J} + 12 H_J^2 \right)\phi_J + 
	\frac{\dd V_J}{\dd \phi_J}
	&= 0, 
	\label{eq:eom1_J} \\
	3M_\text{P}^2H_J^2 + \xi \left( 3H_J^2\phi_J^2 + 6H_J\phi_J \frac{\dd \phi_J}{\dd t_J} \right) 
	&= \frac{1}{2}\left(\frac{\dd \phi_J}{\dd t_J}\right)^2 + V_J, 
	\label{eq:eom2_J} \\
	\left(M_\text{P}^2 + \xi \phi_J^2\right)\left(2\frac{\dd H_J}{\dd t_J} + 3H_J^2\right)
	+ 2\xi\left[\phi_J \frac{\dd^2\phi_J}{\dd t_J^2} + \left(\frac{\dd\phi_J}{\dd t_J}\right)^2 + 2H_J\phi_J\frac{\dd\phi_J}{\dd t_J}\right]
	&= -\frac{1}{2}\left(\frac{\dd\phi_J}{\dd t_J}\right)^2 + V_J,
	\label{eq:eom3_J}
\end{align}
where $\dd t_J \equiv N_J \dd t = \dd t_E/\Omega$ and $H_J = (1/a_J)(\dd a_J/\dd t_J)$.
It is also instructive to
rewrite Eq.~\eqref{eq:eom1_J} as
\begin{align}
&\frac{\dd^2\phi_J}{\dd t_J^2} + 3H_J\frac{\dd\phi_J}{\dd t_J} + m_{J\mathrm{eff}}^2 \phi_J
= 0, \\
&m_{J\mathrm{eff}}^2
= \frac{1}{\phi_J} \frac{\dd V_J }{\dd \phi_J}
+ \frac{\xi (1 + 6\xi)((\dd\phi_J/\dd t_J)^2 - (\dd V_J / \dd \phi_J) \phi_J) 
+ \xi (( \dd V_J / \dd \phi_J )\phi_J - 4V_J )}{M_\text{P}^2 + \xi (1 + 6\xi )\phi_J^2},
\label{eq:phi_tEOM_2}
\end{align}
where we have eliminated $\dd H_J/\dd t_J$ and $H_J^2$ by using Eqs.~\eqref{eq:eom2_J} and~\eqref{eq:eom3_J}.
For the quartic potential, it is further simplified as
\begin{align}
m_{J\mathrm{eff}}^2
&= 
\frac{\xi (1 + 6\xi)(\dd \phi_J/\dd t_J)^2 + \lambda M_\text{P}^2 \phi_J^2}{M_\text{P}^2 + \xi (1 + 6\xi )\phi_J^2}.
\label{eq:app:meff_V4}
\end{align}
The dynamics of the inflaton in the Jordan frame is discussed in detail in App.~\ref{app:bg}.

%%%%%%%%%%%%%%%%%%%%%%%%%%%%%%%%%%%%%%%%%%%%%%%%%%
\subsection{Appearance of spike-like feature} 
\label{sec:mapping}
%%%%%%%%%%%%%%%%%%%%%%%%%%%%%%%%%%%%%%%%%%%%%%%%%%

From now on we concentrate on the inflaton oscillation regime with $M_\text{P}/\xi \lesssim \Phi \lesssim M_\text{P}$
(or equivalently, $M_\text{P}/\xi \lesssim \Phi_J \lesssim M_\text{P}/\sqrt{\xi}$ with $\Phi_J$ being 
the oscillation amplitude of $\phi_J$).
In this subsection we show that there appear two typical mass scales (or inverse time scales) in the inflaton dynamics,
which will be important when discussing the preheating in the next section.
Obviously, one is the inflaton oscillation scale, given by
\begin{align}
	m_{\rm osc} = \Delta t_{\rm osc}^{-1} \sim \sqrt{\frac{\lambda}{3}} \frac{M_\text{P}}{\xi}.
\end{align}
There is another short time scale, during which the inflaton passes through the region $\sqrt{3/2}|\phi| \lesssim M_\text{P}/\xi$.
We call it as the ``spike'' scale for the reason discussed below.
It is estimated as
\begin{align}
	m_{\rm sp} = \Delta t_{\rm sp}^{-1} \sim \sqrt\lambda \Phi \sim \frac{\sqrt \lambda \xi \Phi_J^2}{M_\text{P}}.
\end{align}
Note that this time scale is extremely short just after inflation:
\begin{align}
	m_{\rm sp} \sim \sqrt{\lambda} M_\text{P},
\end{align}
where we have substituted $\Phi \sim M_\text{P}$ or $\Phi_J \sim M_\text{P}/\sqrt\xi$ just after inflation.
This fact has a strong impact on the particle production at the first stage of preheating as we will see later.

\begin{figure}[t]
\begin{center}
\includegraphics[scale=1]{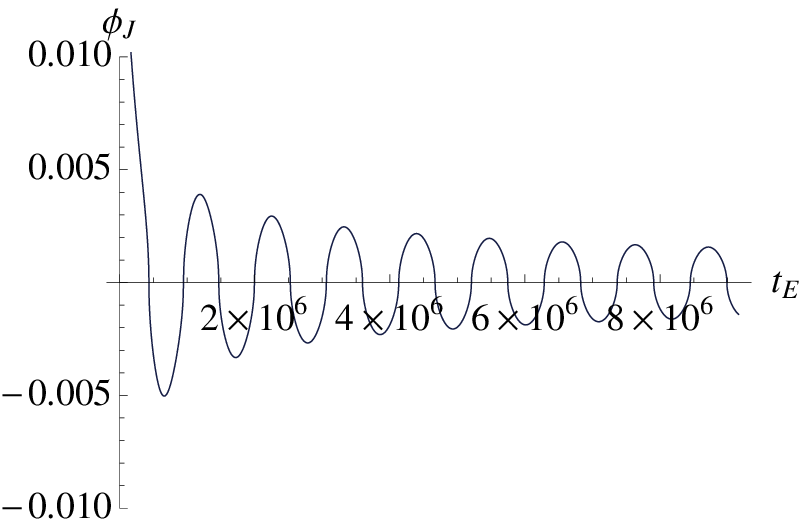}
\includegraphics[scale=1]{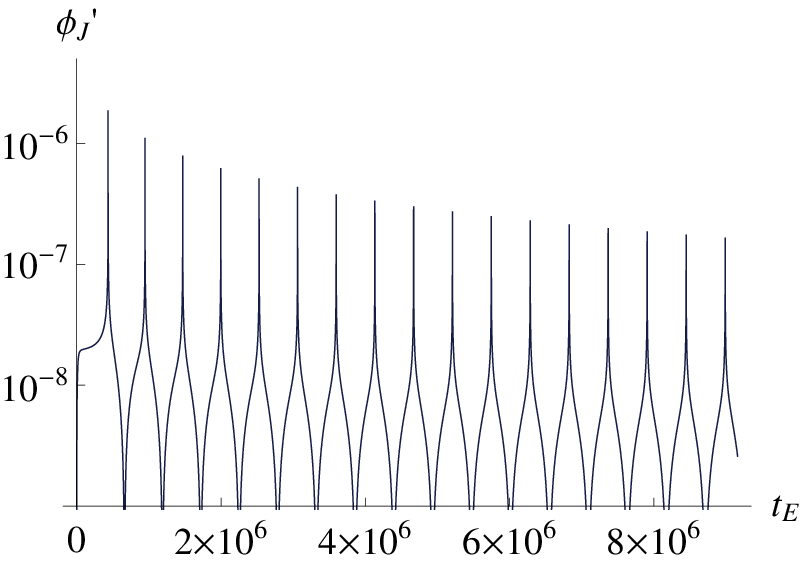}
\caption{ \small
Time evolution of
$\phi_J$ (left) and $|\dot{\phi}_J|$ (right) at the beginning of Phase 1.
Parameters are taken to be $\lambda = 0.01$ and $\xi = 10^4$,
and the initial conditions are $\phi_{J{\rm ini}} = 2M_\text{P}/\sqrt{\xi}$ and $\dot{\phi}_{J{\rm ini}} = 0$.
We take $M_\text{P} = 1$ in this plot.
}
\label{fig:phiJ_dphiJ}
\end{center}
\end{figure}

\begin{figure}[t]
\begin{center}
\includegraphics[scale=1]{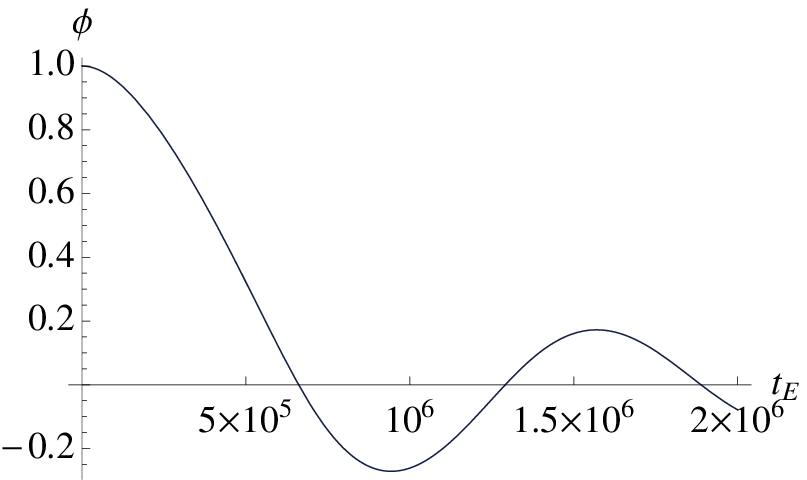}
\includegraphics[scale=1]{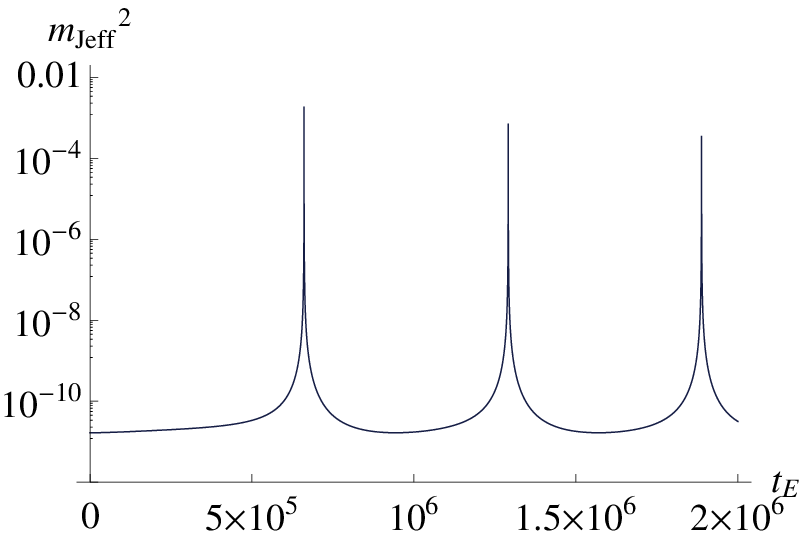}
\caption{ \small
	Time evolution of $\phi$ and $m_{J{\rm eff}}^2$
	around the first few zero-crossings.
	The parameter values are $\lambda = 0.01$ and $\xi = 10^4$,
	while the initial conditions are $\phi_{{\rm ini}} = M_\text{P}$ 
	and $\dot{\phi}_{{\rm ini}} = 0$.  
	We can clearly see that $m_{J{\rm eff}}^2$ blows up around the zero-crossings,
	thus representing a spike-like feature.
	We take $M_\text{P} = 1$ in this plot.
}
\label{fig:phi_mJ2}
\end{center}
\end{figure}

Now we see why the spike scale appears in the dynamics of $\phi_J$ and $\phi$.
Let us first consider the dynamics of $\phi_J$.
In this case, the spike time scale appears as a change of the kinetic term of $\phi_J$.
In fact, from Eq.~\eqref{eq:eom3_J}, one sees that
\begin{align}
	\frac{\dd H_J}{\dd t_J} = -\frac{\xi \phi_J}{M_\text{P}^2 + \xi \phi_J^2} \frac{\dd^2 \phi_J}{\dd t_J^2} + (\mathrm{other~terms}),
\end{align}
where the rest part depends only on $\phi_J$, $\dd\phi_J/\dd t_J$ and $H_J$.
It means that there is a kinetic mixing between the scalar component of the metric and $\phi_J$
due to the nonminimal coupling.
By substituting this into Eq.~\eqref{eq:eom1_J}, one can see that the mixing term dominates
over the original kinetic term of $\phi_J$
(\textit{i.e.} $g^{\mu\nu}_J\partial_\mu \phi_J \partial_\nu \phi_J/2$)
for $\lvert \phi_J \rvert \gg M_\text{P}/\xi$.
Namely, the kinetic term of $\phi_J$ changes for 
$\lvert \phi_J \rvert \gg M_\text{P}/\xi$ and $\lvert \phi_J \rvert \ll M_\text{P}/\xi$.
The original kinetic term is dominant only for the time scale $\Delta t_\mathrm{sp}$,
and this is the physical reason why the spike scale appears in the dynamics of $\phi_J$.
In order to see how the behavior of $\phi_J$ actually changes during this time interval $\Delta t_\mathrm{sp}$,
it is helpful to solve Eq.~\eqref{eq:phi_tEOM_2} by neglecting the Hubble friction term.
As calculated in detail in App.~\ref{app:bg}, the solution is approximated as
\begin{align}
	\phi_J \sim 
		\pm \Phi_J \sqrt{|\cos(m_{\rm osc} t)|}  ~~~~{\rm for}~~~~|\phi| \gtrsim M_\text{P}/\xi.
\end{align}
Hence its time derivative tends to diverge as the inflaton approaches to the origin $\phi_J \to 0$.
The typical magnitude of its time derivative is
\begin{align}
	\dot\phi_J \sim \begin{cases}
		\displaystyle \sqrt{\lambda} \Phi_J^2 \sim \frac{\sqrt{\lambda}M_\text{P}}{\xi}\Phi
		&\displaystyle {\rm for}~~|\phi| \ll \frac{M_\text{P}}{\xi}, \\
		\displaystyle \frac{\sqrt \lambda M_\text{P}}{\xi}\Phi_J 
		\sim \sqrt{\frac{\lambda M_\text{P}^3 \Phi}{\xi^3}}
		&\displaystyle {\rm for}~~|\phi| \gg \frac{M_\text{P}}{\xi}.
	\end{cases}
\end{align}
Thus this quantity around the origin is enhanced by a factor of $\sqrt{\xi\Phi/M_\text{P}}~(\gg 1)$ 
compared with that far from the origin.
As another example, we consider the effective inflaton mass in the Jordan frame,
$m_{J {\rm eff}}^2$ ($\sim \ddot{\phi}_J/\phi_J$), defined in Eq.~(\ref{eq:app:meff_V4}). It behaves as
\begin{align}
	m_{J {\rm eff}}^2 \sim \begin{cases}
		\displaystyle \frac{\lambda\xi^2 \Phi_J^4}{M_\text{P}^2} \sim \lambda \Phi^2
		&\displaystyle {\rm for}~~|\phi| \ll \frac{M_\text{P}}{\xi},\\[.8em]
		\displaystyle \frac{\lambda M_\text{P}^2}{\xi^2} &\displaystyle {\rm for}~~|\phi| \gg \frac{M_\text{P}}{\xi}.
	\end{cases}
	\label{mJeff}
\end{align}
Again it is enhanced at around the origin by $\xi^2\Phi^2/M_\text{P}^2~(\gg 1)$.
In order to confirm these statements, we numerically follow the time evolution of 
$\phi_J$ and $\dot\phi_J$ in Fig.~\ref{fig:phiJ_dphiJ},
and that of $m_{J\mathrm{eff}}^2$ in Fig.~\ref{fig:phi_mJ2}, respectively.
We can see the ``spike''-like feature in $\dot\phi_J$ and $m_{J\mathrm{eff}}^2$
at around when $\phi$ crosses the origin,
and the height and timescale of the spike-like feature are consistent with our estimation. 
This is the reason why we called $\Delta t_{\rm sp}$ as the spike time scale.
Note that the enhancement of $m_{J\mathrm{eff}}^2$ is much more violent than that of $\dot\phi_J$.

%%%%%%%%%%%%%%%%
%%%%%%%%%%%%%%%%

Therefore we expect that any function consisting of $\dot\phi_J, \ddot\phi_J,\dots$ also exhibits 
a spike-like feature.
This fact has significant impacts on the preheating after inflation 
since typically the effective masses of the coupled particles
are actually functions of them, not of $\phi$.
This is because the theory is usually defined in a simple form in the Jordan frame,
not in the Einstein frame.
It is well known that a sudden change of the effective mass terms causes particle production.
Thus, if there are particles that couple to, \textit{e.g.} $m_{J{\rm eff}}^2$, they are excited due to the spike-like feature.
Quite interestingly, if the inflaton is a complex scalar field, there are indeed such particles and
it provides a main channel to transfer the energy density of the inflaton to other particles.
As specific examples, we consider the inflaton with a global/gauged U(1) charge in the next sections
(Secs.~\ref{sec:global_U1} and~\ref{sec:gauged_U1}, respectively).
For the global U(1) case, the U(1) partner of the inflaton is violently produced
due to the spike-like feature of $m_{J{\rm eff}}^2$.
For the gauged U(1) case, the longitudinal mode of the gauge boson plays the same role.

%%%%%%%%%%%%%%%%
%%%%%%%%%%%%%%%%

\begin{figure}[t]
\begin{center}
\includegraphics[scale=1]{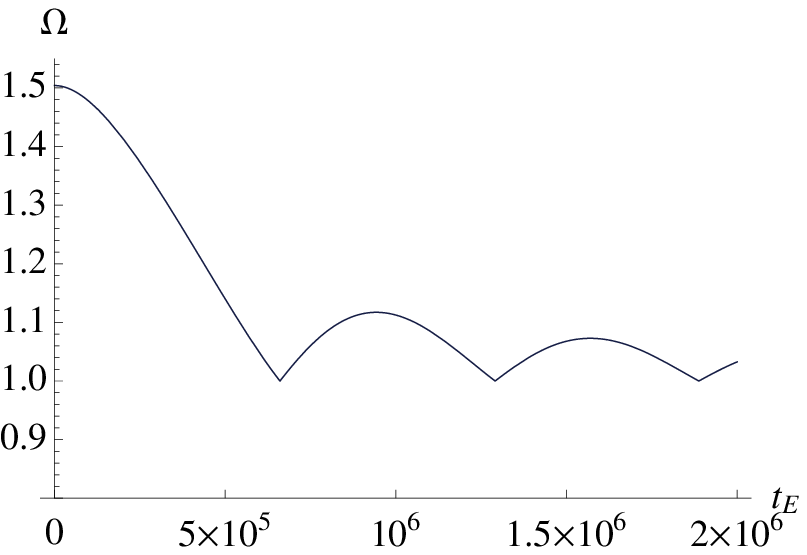}
\includegraphics[scale=1]{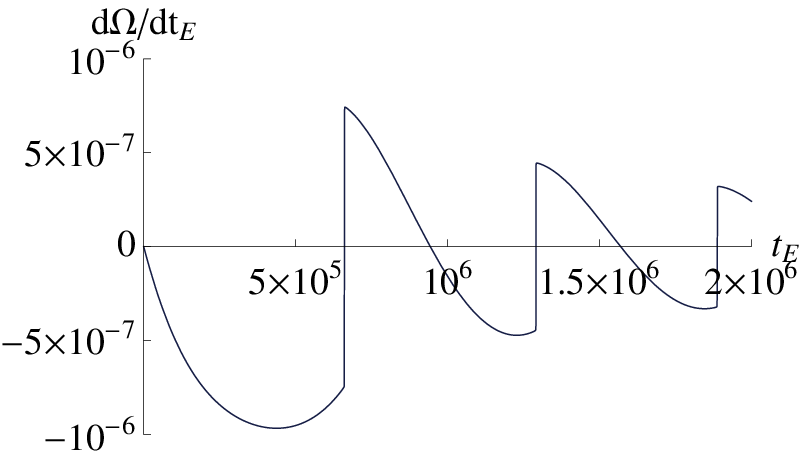}
\caption{ \small
Time evolution of $\Omega$ (left) and $\dot{\Omega}$ (right).
The parameters are taken to be $\lambda = 0.01$ and $\xi = 10^4$,
and the initial conditions are $\phi_{\rm ini} = M_\text{P}$ and $\dot{\phi}_{\rm ini} = 0$.
We can see that $\dot{\Omega}$ takes a typical value $\sim \sqrt{\lambda}\Phi / \xi$
throughout the oscillation.
We take $M_\text{P} = 1$ in this plot.
}
\label{fig:Omega_dOmega}
\end{center}
\end{figure}

\begin{figure}[t]
\begin{center}
\includegraphics[scale=1]{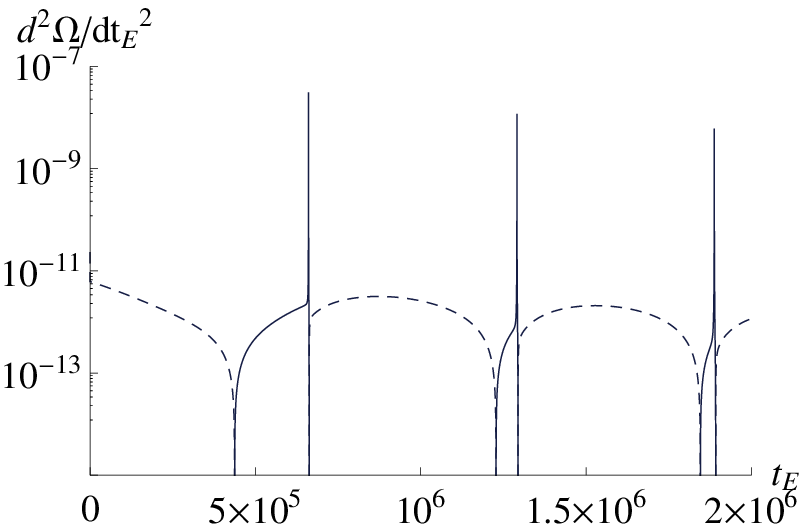}
\includegraphics[scale=1]{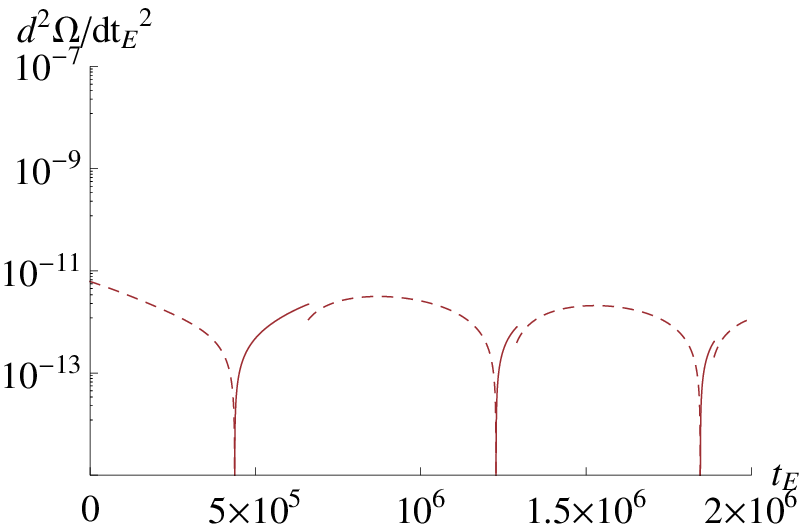}
\caption{ \small
Time evolution of $\ddot{\Omega}$ with the exact system (left)
and the approximated system (right).
In both panels, the solid and dashed line show $\ddot{\Omega}$ and
$- \ddot{\Omega}$, respectively.
The parameters and initial conditions are the same as in Fig.~\ref{fig:Omega_dOmega}.
We can see that $\ddot{\Omega}$ takes $\sim \lambda\Phi^2 / \xi$
only when $\phi$ crosses the origin.
For the jump in the value of $\ddot{\Omega}$ in the right panel, 
see the footnote~\ref{fn:jump} in the main text.
We take $M_\text{P} = 1$ in this plot.
}
\label{fig:ddOmega}
\end{center}
\end{figure}

\begin{figure}[t]
\begin{center}
\includegraphics[scale=1]{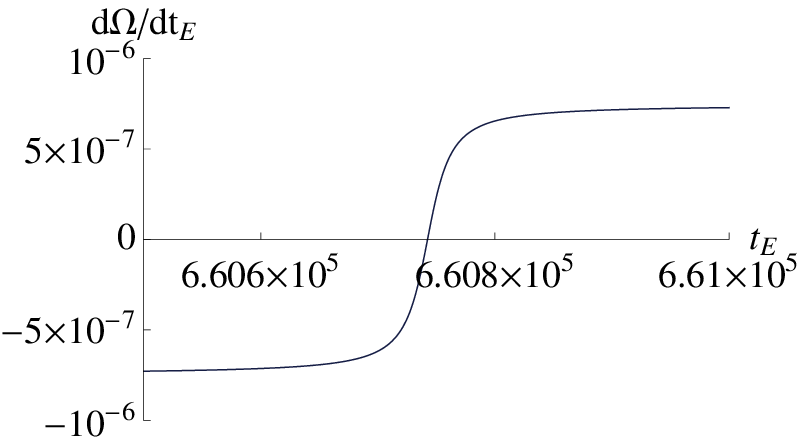}
\includegraphics[scale=1]{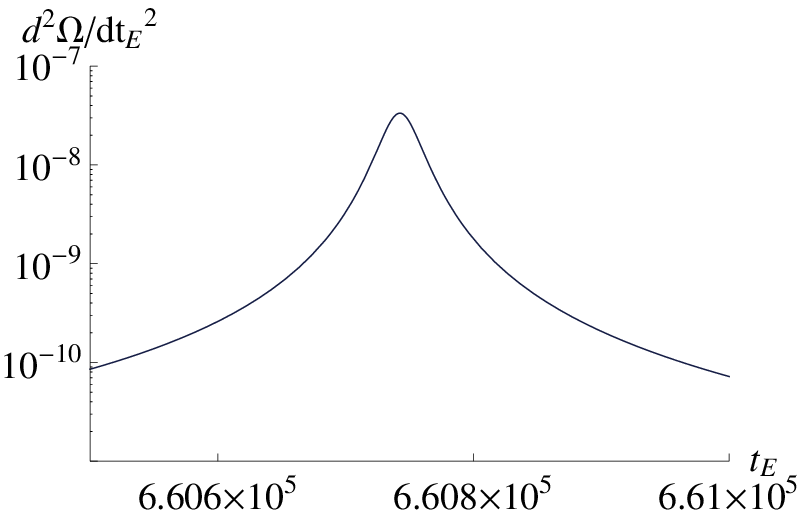}
\caption{ \small
Magnifications of Fig.~\ref{fig:Omega_dOmega} and \ref{fig:ddOmega}
around the first zero-crossing of $\phi$.
}
\label{fig:dOmega_ddOmega_blowup}
\end{center}
\end{figure}

As another example, we show the behavior of the conformal factor $\Omega$,
which will also become important for the discussion of 
the preheating of a minimally coupled scalar field in Sec.~\ref{sec:real}.
The first derivative of the conformal factor has roughly the same orders-of-magnitude values
for $\lvert \phi \rvert \ll M_\text{P}/\xi$ and $\lvert \phi \rvert \gg M_\text{P}/\xi$:
\begin{align}
\dot{\Omega} \simeq \frac{\xi\phi_J\dot\phi_J}{M_\text{P}^2}
& \sim \frac{\sqrt{\lambda}\Phi_J^2}{M_\text{P}} \sim\frac{\sqrt{\lambda}\Phi}{\xi}.
\end{align}
However, the second derivative shows a clear difference:
\begin{align}
\ddot{\Omega} \simeq \frac{\xi (\phi_J\ddot\phi_J + \dot\phi_J^2)}{M_\text{P}^2}
&\sim
\begin{cases}
\displaystyle \frac{\lambda \xi \Phi_J^4}{M_\text{P}^2} \sim \frac{\lambda\Phi^2}{\xi} 
~~~~~~
&\mathrm{for}
~~~
|\phi| \ll \displaystyle\frac{M_\text{P}}{\xi} ,
\\[.8em]
\displaystyle\frac{\lambda \Phi_J^2}{\xi} \sim \frac{\lambda M_\text{P} \Phi}{\xi^2}
~~~~~~
&\mathrm{for}
~~~
|\phi| \gg \displaystyle\frac{M_\text{P}}{\xi}. 
\end{cases}
\label{ddotOmega}
\end{align}
The order of magnitude of $\ddot{\Omega}$ is different by a large factor $\mathcal{O}(\xi \Phi/M_\text{P})$
for $\lvert \phi \rvert \ll M_\text{P}/\xi$ and $\lvert \phi \rvert \gg M_\text{P}/\xi$.
We numerically follow the time evolution of 
$\Omega$, $\dot{\Omega}$ and $\ddot{\Omega}$
in Figs.~\ref{fig:Omega_dOmega}--\ref{fig:dOmega_ddOmega_blowup}.
\textcolor{dark_blue}{The blue line} corresponds to the exact solution,
while \textcolor{dark_red}{the red line}
does to the approximated one using
\begin{align}
	\frac{\lvert\phi\rvert}{M_\text{P}} \simeq \sqrt{\frac{3}{2}}\ln \left(1+\frac{\xi \phi_J^2}{M_\text{P}^2}\right), ~~
	V(\phi) \simeq \frac{\lambda M_\text{P}^4}{4\xi^2}\left[1 - \exp\left(-\sqrt{\frac{2}{3}}\frac{\lvert\phi\rvert}{M_\text{P}}\right) \right]^2,
	\label{eq:bad_approx}
\end{align}
in the whole region, as often used in the literature.
For $\Omega$ and $\dot{\Omega}$ we plot only exact solutions
because the approximated ones show no visible difference. 
However, $\ddot{\Omega}$ shows a clear difference when the inflaton crosses the origin. 
In fact, $\ddot{\Omega}$ shows a spike at around the zero-crossings in the exact case,
while such a feature is not captured by the approximated one\footnote{
\label{fn:jump}
The value of $\ddot{\Omega}$ jumps when the inflaton crosses the origin 
as long as one uses the approximation~\eqref{eq:bad_approx}.
This is because we have to take the absolute value to keep the $\mathbb{Z}_2$ symmetry in Eq.~\eqref{eq:bad_approx}.
}.

%%%%%%%%%%%%%%%%
%%
%%
%%%%%%%%%%%%%%%%

Next we consider the dynamics of $\phi$.
In this case, the spike scale appears as a sudden change in the shape of the potential.
For $\Phi \gg M_\text{P}/\xi$, the equation of motion reads
\begin{align}
	\ddot\phi \simeq 
	\begin{cases}
	-\lambda\phi^3 &\displaystyle {\rm for}~~|\phi| \ll \frac{M_\text{P}}{\xi}, 
	\\[.8em]
	-m_{\rm osc}^2 \phi &\displaystyle {\rm for}~~|\phi| \gg \frac{M_\text{P}}{\xi}.
	\end{cases}
\end{align}
Therefore the effective mass of $\phi$ in the Einstein frame, $m_{\rm eff}^2$, suddenly drops to zero
in the short time interval $\Delta t_{\rm sp}$ as
\begin{align}
	m_{\rm eff}^2 \equiv \frac{\partial^2 V}{\partial \phi^2} \sim 
	\begin{cases}
		0  &\displaystyle {\rm for}~~|\phi| \ll \frac{M_\text{P}}{\xi}, \\[.8em]
		\displaystyle m_{\rm osc}^2 = \frac{\lambda M_\text{P}^2}{3\xi^2} &\displaystyle {\rm for}~~|\phi| \gg\frac{M_\text{P}}{\xi}.
	\end{cases}
\end{align}
This can also be interpreted as a spike and is related with the inflaton self-production 
as studied in Sec.~\ref{sec:real},\footnote{
	The existence of this spike scale in the Higgs self-production 
	was briefly mentioned in Ref.~\cite{Bezrukov:2008ut}.
}
although the effect is rather mild compared with the case of $m_{J {\rm eff}}^2$.
These considerations clearly show that the efficiency of preheating or particle production strongly depends on
whether light particles couple to $\phi$ or $\phi_J$ (and their derivatives).
Usually, as in the Higgs inflation, the Lagrangian is defined in the Jordan frame in which all terms have simple forms.
In such a case, a violent particle production is inevitable as we will see in the next section.

It may be instructive to summarize the relation between $\phi_J$ and $\phi$ and their time derivatives here.
Clearly, we have $\phi\simeq \phi_J$ for $\lvert \phi \rvert \ll M_\text{P}/\xi$.
However, their time derivatives do not always coincide with each other. 
Actually we obtain the relation between $\phi$ and $\phi_J$ for $\lvert \phi \rvert \ll M_\text{P}/\xi$ by taking the time derivative of Eq.~\eqref{eq:mapping} as
\begin{align}
	\phi &\simeq \phi_J, \\
	\dot{\phi} &\simeq \dot{\phi}_J, \\
	\ddot{\phi} &\simeq \ddot{\phi}_J - \frac{\xi(1-6\xi)\dot{\phi}_J^2}{M_\text{P}^2}\phi_J.
	\label{eq:phi_J_second}
\end{align}
The last equation~\eqref{eq:phi_J_second} is the most important.
We know that $\ddot{\phi} \sim \lambda \phi^3 \sim \lambda \phi_J^3$ and 
$\dot{\phi} \sim \dot{\phi}_J \sim \sqrt{\lambda}M_\text{P}^2/\xi$
at around the first zero-crossing of the inflaton.
In Eq.~\eqref{eq:phi_J_second}, the second term in the R.H.S. 
is larger than the L.H.S. at least by ${\mathcal O}(\xi^2)$
for $\lvert \phi \rvert \ll M_\text{P}/\xi$.
Hence, the first and second terms in the R.H.S. in Eq.~\eqref{eq:phi_J_second} must be almost cancelled out:
\begin{align}
	\ddot{\phi}_J \simeq \frac{\xi(1-6\xi)\dot{\phi}_J^2}{M_\text{P}^2} \phi_J \sim  \frac{\lambda\xi^2\Phi_J^4}{M_\text{P}^2} \phi_J
	~~~
	\mathrm{for}
	~~~
	\lvert \phi \rvert \ll \frac{M_\text{P}}{\xi}.
	\label{eq:cancel}
\end{align}
It suggests that the mass scale of $\phi_J$ blows up 
for $\lvert \phi \rvert \ll M_\text{P}/\xi$.
It corresponds to the blow-up of $m_{J\mathrm{eff}}^2$ at around the origin as we have already seen before.
Note that because the mass scale blows up only for an extremely short interval,
the condition $\ddot{\phi} ~ (\sim \lambda \phi^3) \ll \ddot{\phi}_J$
for $\lvert \phi \rvert \ll M_\text{P}/\xi$
is compatible with $\phi \simeq \phi_J$ and $\dot{\phi} \simeq \dot{\phi}_J$.
In fact, 
although the mass scale blows up,
the  growth of the first derivative is at most comparable 
\begin{align}
\ddot{\phi}_J \Delta t_{\rm sp} \lesssim \frac{\lambda \xi\Phi_J^4}{M_\text{P}}\frac{M_\text{P}}{\sqrt{\lambda} \xi\Phi_J^2}
\sim \dot{\phi}_J,
\end{align}
and thus the velocities $\dot{\phi}$ and $\dot{\phi}_J$ remain roughly the same
before and after the spike.

%%%%%%%%%%%%%%%%%%%%%%%%%%%%%%%%%%%%%%%%%%%%%%%%%%
\section{Preheating}
\label{sec_pp}
\setcounter{equation}{0}
%%%%%%%%%%%%%%%%%%%%%%%%%%%%%%%%%%%%%%%%%%%%%%%%%%

As we saw in the previous section, 
functions such as $m_{J\mathrm{eff}}^2$ and $\ddot{\Omega}$ show spike-like behavior at around the origin.
This spike works as external force for particles, by which particle production is triggered.
In this section, we study this process in detail.
We briefly summarize basic ingredients of particle production by such spike-like external force
in Sec.~\ref{sec:pp_spike} (see App.~\ref{app:pp} for more details).
Then, we analyze particle production in real, global U(1), and gauged U(1) inflaton cases
in Secs.~\ref{sec:real}, \ref{sec:global_U1} and~\ref{sec:gauged_U1}, respectively.

We are interested in the very first stage of the preheating and
we do not discuss processes after that, such as parametric resonance, decay and annihilation of produced particles
and thermalization, since these processes strongly depend on the properties of coupled particles.
On the other hand, we can discuss the first stage of the preheating almost model independently.

We proceed in the Einstein frame in this section.
The equivalence between the Einstein and Jordan frames is discussed in App.~\ref{app:frame}.

%%%%%%%%%%%%%%%%%%%%%%%%%%%%%%%%%%%%%%%%%%%%%%%%%%
\subsection{Particle production by spike-like external force}
\label{sec:pp_spike}
%%%%%%%%%%%%%%%%%%%%%%%%%%%%%%%%%%%%%%%%%%%%%%%%%%

In this subsection, we summarize basic ingredients to study particle production 
by the spike-like external force we encountered in the previous section, 
\textit{i.e.}~$m_{J\mathrm{eff}}^2$ and $\ddot{\Omega}$.
We consider a real canonical scalar field $\chi$ which satisfies the following equation of motion:
\begin{align}
	\ddot{\chi}_{k} + \omega_{k}^2(t)\chi_k = 0,
	~~
	\omega_{k}^2 = k^2 + m_\chi^2(t),
	\label{eq:basic_eom}
\end{align}
where $m_{\chi}$ is a time-dependent mass
which causes particle production.
Here we neglect the scale factor since the mass scale of the 
spike-like feature $m_\mathrm{sp}$ satisfies $m_\mathrm{sp} \gg H$.
We will find the explicit form of $m_\chi^2$ in terms of $\phi_J$ or $\phi$ 
in the following subsections.
Once we find it, 
it is straightforward to calculate the production rate of $\chi$ using the Bogoliubov transformation.
The time evolution of the Bogoliubov coefficients $\alpha_k(t)$ and $\beta_k(t)$ is described by
\begin{align}
\dot{\alpha}_k(t)
&= \frac{\dot{\omega}_k}{2\omega_k}e^{2i\int_{-\infty}^t \dd t'\omega_k(t')}\beta_k(t),
\;\;\;\;
\dot{\beta}_k(t)
= \frac{\dot{\omega}_k}{2\omega_k}e^{-2i\int_{-\infty}^t \dd t'\omega_k(t')}\alpha_k(t).
\label{eq:EOM_alphabeta}
\end{align}
The occupation number $f_\chi$ is given by
\begin{align}
	f_\chi(t, k) 
	= \left\lvert \beta_{k}(t)\right\rvert^2,
\end{align}
with which we can express the produced number/energy density of $\chi$ as
\begin{align}
n_\chi(t)
&= \int \frac{\dd^3k}{(2\pi)^3} \;
f_\chi(t,k),
\;\;\;\;
\rho_\chi(t)
= \int \frac{\dd^3k}{(2\pi)^3} \;
\omega_k(t) f_\chi(t,k).
\end{align}
The vacuum initial conditions are
\begin{align}
	\alpha_k(t\rightarrow -\infty) = 1,
	~~
	\beta_k(t\rightarrow -\infty) = 0.
\end{align}
For more details, see App.~\ref{app:pp}.
Thus, we only have to numerically solve Eq.~\eqref{eq:EOM_alphabeta} 
to obtain the produced amount of $\chi$.
Indeed, this is what we do in the subsequent sections.
However, it should be still of some help 
to understand qualitative features of particle production.
Therefore, in the rest of this subsection, 
we analytically estimate the produced amount of $\chi$
caused by the spike-like external force.
We will compare it with numerical results in the following subsections.

Since we are interested in particle production by spike-like external force,
we assume that $m_\chi^2$ takes the following form:
\begin{align}
	m_\chi^2(t)
	&=
	\bar{m}_\chi^2 ~ \mathrm{sp}(m_\mathrm{sp}t),
\end{align}
where $\bar{m}_{\chi}$ is the over-all normalization of the time-dependent mass
and $\mathrm{sp}(m_\mathrm{sp}t)$ denotes a ``spike function'' whose maximal value is unity.
We require $\bar{m}_\chi \lesssim m_\mathrm{sp}$ and 
$\mathrm{sp}(m_\mathrm{sp}t) \rightarrow 0$ for $\lvert m_\mathrm{sp}t\rvert \gg 1$.
In the case of our interest, the mass scale of the spike is $m_\mathrm{sp} \sim \sqrt{\lambda}M_\text{P}$.
In realistic cases, the spike function is expressed by $\phi_J$ in a complicated manner,
and hence it is difficult to analytically solve Eq.~\eqref{eq:EOM_alphabeta}.
Instead, here we take a simpler form for $\mathrm{sp}(m_\mathrm{sp}t)$,
and try to capture generic features of particle production caused by the spike-like force.
As examples, we take $\mathrm{sp}(m_\mathrm{sp}t)$ 
to be trigonometric or gaussian~(Eq.~\eqref{eq:sp_simple}).
For these cases, we can solve Eq.~\eqref{eq:EOM_alphabeta} to obtain\footnote{
At $k \simeq m_{\rm sp}$, the values of $f_\chi$ in Eq.~(\ref{eq:pp_f_summary}) 
and $d\rho_\chi/d\ln k$ in Eq.~(\ref{eq:pp_drho_summary}) are suppressed by a factor of 
${\mathcal O}(0.1)$ compared to the values with $k = m_{\rm sp}$ substituted in these equations.
See App.~\ref{app:pp} for details.
\label{fn:f_drho}
}
\begin{align}
f_\chi
&\sim
\frac{\bar{m}_\chi^4}{m_{\rm sp}^2k^2}
~~~
\mathrm{for}
~
k \ll m_{\rm sp},
\label{eq:pp_f_summary}
\end{align}
and there is a cut-off for $k \gg m_\mathrm{sp}$ for both cases.
Although the shape of the cut-off depends on the explicit form of $\mathrm{sp}(m_\mathrm{sp}t)$,
we may expect that the momentum distribution is generically 
expressed as Eq.~\eqref{eq:pp_f_summary} plus a rapid fall-off\footnote{
	Faster than $k^{-4}$ at least.
} for $k \gg m_\mathrm{sp}$
for the particle production caused by the spike-like external force.
The resulting energy spectrum is
\begin{align}
\frac{\dd\rho_\chi}{\dd\ln k}
&\sim 
\frac{1}{2\pi^2} \frac{\bar{m}_\chi^4k^2}{m_{\rm sp}^2}
~~~
\mathrm{for}
~
k \ll m_{\rm sp},
\label{eq:pp_drho_summary}
\end{align}
and again there is a cut-off for $k \gg m_\mathrm{sp}$.
Therefore, the energy density is dominated by particles with 
$k \sim m_{\rm sp}$, and the total energy density after one inflaton oscillation is estimated as
\begin{align}
\rho_\chi
&\sim 
\frac{\bar{m}_\chi^4}{4\pi^2}.
\label{eq:pp_rho_summary}
\end{align}
Thus, if $\bar{m}_\chi$ is large enough, 
almost all the energy of inflaton might be transferred to $\chi$.
For more details, see App.~\ref{app:pp}.

%%%%%%%%%%%%%%%%%%%%%%%%%%%%%%%%%%%%%%%%%%%%%%%%%%
\subsection{Real scalar inflaton}
\label{sec:real}
%%%%%%%%%%%%%%%%%%%%%%%%%%%%%%%%%%%%%%%%%%%%%%%%%%

Now we study particle production caused 
by the spike-like feature in several concrete models. 
Here we discuss the case where the inflaton is a real scalar field.
We consider two cases in this subsection:  production of (1) the inflaton particle itself, 
and (2) an additional light scalar field.
For the second case, we assume that the light scalar field has a minimal kinetic term in the Jordan frame,
not in the Einstein frame.

%%%%%%%%%%%%%%%%%%%%%%%%%%%%%%%%%%%%%%%%%%%%%%%%%%
\subsubsection*{Self production}
%%%%%%%%%%%%%%%%%%%%%%%%%%%%%%%%%%%%%%%%%%%%%%%%%%

First we study the self production of the inflaton.
Note that the inflaton particle has a mixing with the scalar components
of the metric because the inflaton has a time-dependent vacuum expectation value (VEV),
and hence we should first solve the mixing.
In the Einstein frame the effect of mixing is not so important for the estimate of particle production, but 
in the Jordan frame taking account of the mixing is essential, as shown in App.~\ref{app:frame_gauge}.
However, it is instructive to consider the mixing even in the Einstein frame and below we study it.

\begin{figure}[t]
\begin{center}
\includegraphics[scale=1]{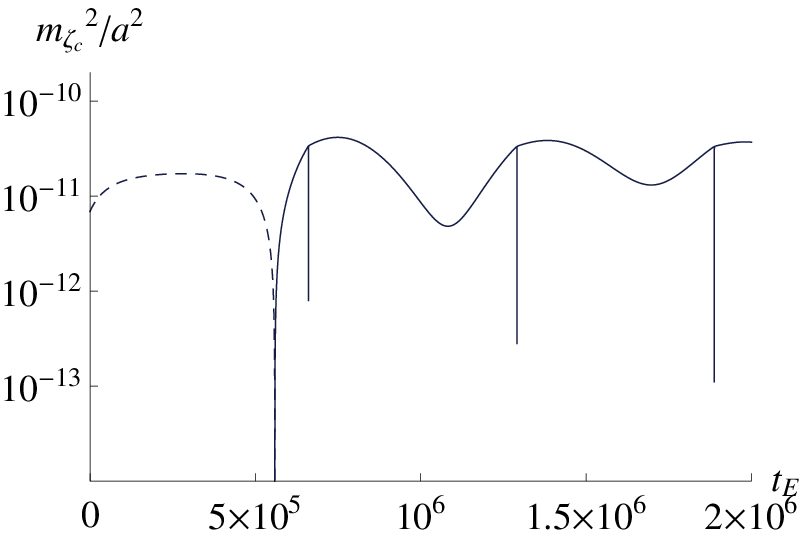}
\includegraphics[scale=1]{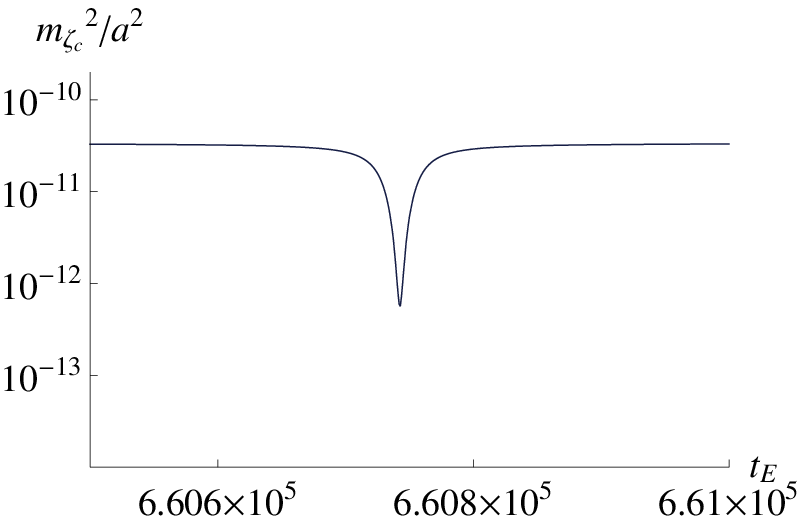}
\caption{ \small
Time evolution of $m_{\zeta_c}^2/a^2$ during the first few zero-crossing of the inflaton.
The parameter values are the same as in Fig.~\ref{fig:phi_mJ2}.
In the left panel, the solid line shows $m_{\zeta_c}^2/a^2$ 
while the dashed line shows $-m_{\zeta_c}^2/a^2$.
The value of $m_{\zeta_c}^2/a^2$ drops from 
$\dd^2V/\dd \bar{\phi}^2 \sim \lambda M_\text{P}^2/\xi^2$ to ${\mathcal O}(H^2)$ 
during the zero-crossings.
The right panel is the magnification of the left panel.
We take $M_\text{P} = 1$ in this plot.
}
\label{fig:mzeta}
\end{center}
\end{figure}

\begin{figure}[t]
\begin{center}
\includegraphics[scale=0.9]{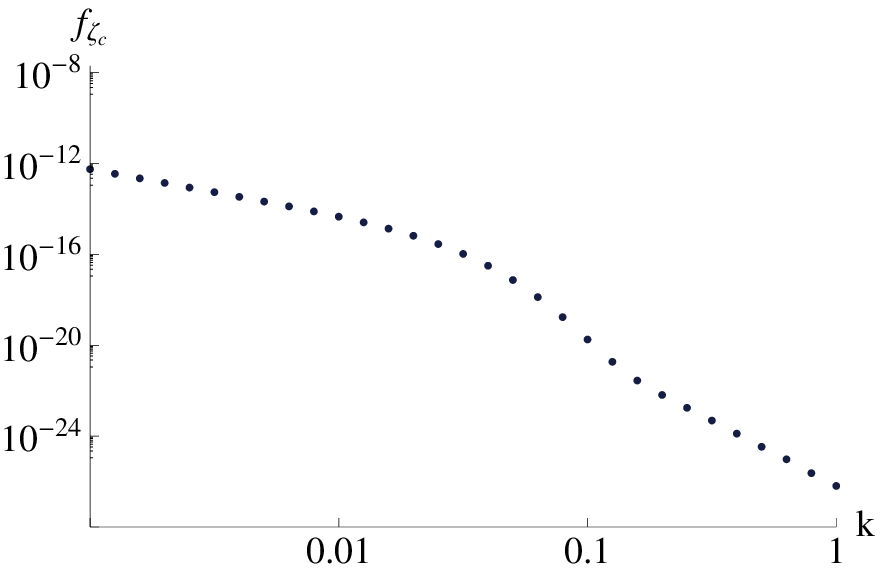}
\includegraphics[scale=0.9]{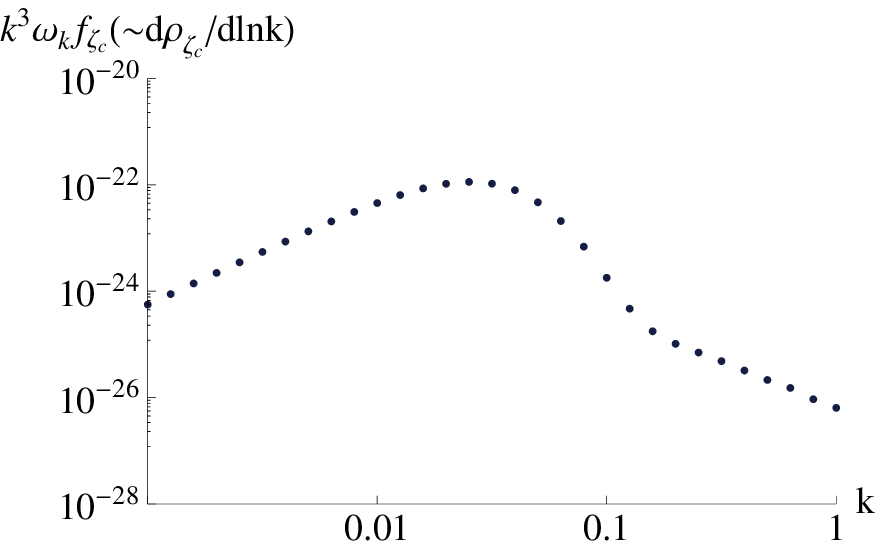}
\caption{ \small
The occupation number $f_{\zeta_c}$ (left) and the energy density spectrum 
$k^3 \omega_k f_{\zeta_c}$ (right) of $\zeta_c$ produced by the first zero-crossing of $\phi$.
The parameters and initial conditions are the same as in the previous figure,
and we take $M_\text{P} = 1$.
In the right panel, one sees that 
the energy density of the produced particles is much smaller than the inflaton energy density 
($\sim 10^{-12}$ in the Planck unit).
}
\label{fig:fzeta_rhozeta}
\end{center}
\end{figure}

We decompose the metric in the Einstein frame using the ADM formalism~\cite{Arnowitt:1962hi}
\begin{align}
\dd s^2
&= -N^2 \dd t^2 + a^2 e^{2\zeta}(\dd x^i + \beta^i \dd t)(\dd x^i + \beta^i \dd t),
\end{align}
where we concentrate only on the scalar components here.
By taking the unitary gauge $\phi(t, \vec{x}) = \bar{\phi}(t)$,
we can solve the mixing and obtain the quadratic action for $\zeta$ as
\begin{align}
S_{\zeta}
&= \int \dd\tau \dd^3x \; a^2
\frac{\bar{\phi}'^2}{2 {\mathcal H}^2}
\left[
\zeta'^2  - (\partial_i \zeta)^2
\right],
\label{eq:pp_zeta_E_conformal}
\end{align}
where the conformal time $\tau$ is defined as $\dd t_E = N \dd t = a \dd \tau$, 
the prime denotes the derivative with respect to $\tau$,
and ${\mathcal H} \equiv aH$.
In order to make $\zeta$ canonical, we define
\begin{align}
\zeta_c
&\equiv F \zeta,
\;\;\;\;
F
\equiv \frac{a\bar{\phi}'}{{\mathcal H}}.
\end{align}
Due to this rescaling, $\zeta_c$ obtains a time-dependent mass term as (see App.~\ref{app:frame})
\begin{align}
m_{\zeta_c}^2
&= 
- \frac{F''}{F}.
\end{align}
By using the background equations of motion, we obtain
\begin{align}
m_{\zeta_c}^2
&= 
a^2\frac{\dd^2V}{\dd\bar{\phi}^2}
- 2{\mathcal H}^2 
- \frac{\bar{\phi}'^2}{2M_\text{P}^2}
- \frac{\bar{\phi}'^4}{2M_\text{P}^4{\mathcal H}^2}
- \frac{(\bar{\phi}'^2)'}{M_\text{P}^2{\mathcal H}}.
\label{eq:mass_zeta}
\end{align}
In this expression of $m_{\zeta_c}^2/a^2$, 
the first term dominates over the others during almost the whole period of oscillation.
However, it becomes $\dd^2V/\dd\bar{\phi}^2 \sim \lambda \bar{\phi}^2$ at around the origin
during a period of $\sim (\sqrt{\lambda}M_\text{P})^{-1}$,
and this sudden change in the potential shape 
causes particle production with momenta 
$\sim \sqrt{\lambda}M_\text{P}$~\cite{Bezrukov:2008ut}.
The left panel of Fig.~\ref{fig:mzeta} shows the time evolution of the mass squared of $\zeta_c$
around the first a few zero-crossings of the inflaton.
The parameter values are taken to be $\lambda = 0.01$ and $\xi = 10^4$, 
and the initial condition of the inflaton is set to be $\phi = M_\text{P}$ with vanishing velocity.
The solid line shows $m_{\zeta_c}^2/a^2$, while the dashed line corresponds to $-m_{\zeta_c}^2/a^2$.
The corresponding evolution of the inflaton is already shown in Fig.~\ref{fig:phi_mJ2},
and one sees that, after the inflaton starts oscillating, $m_{\zeta_c}^2/a^2$
drops from $\dd^2V/\dd \bar{\phi}^2 \sim \lambda M_\text{P}^2/(3\xi^2)$ to the order of $H^2$ 
when the inflaton crosses the origin.
Thus in this case the height of the spike, $\bar m_{\chi}$ in Eq.~(\ref{eq:pp_drho_summary}), is given by
$\bar m_\chi \sim m_{\rm osc} \sim \sqrt{\lambda}M_\text{P}/\xi$.
The energy density of inflaton fluctuation after the first zero-crossing of the inflaton is then 
roughly estimated as
\begin{align}
	\rho_\zeta \sim \frac{\lambda^2 M_\text{P}^4}{\xi^4}.  \label{rho_zeta}
\end{align}
To confirm this, we numerically calculated 
the produced amount of $\zeta_c$ during the first zero-crossing of the inflaton
with the same parameter point as in Fig.~\ref{fig:mzeta}.
We solved Eq.~\eqref{eq:EOM_alphabeta} with Eq.~\eqref{eq:mass_zeta}
neglecting the overall scale factor,
and evaluated the Bogoliubov coefficients well after the spike.\footnote{
We evaluated the Bogoliubov coefficients
at the time after the first spike by $\Delta t \simeq 5\times10^4M_\text{P}^{-1}$,
which is well after the spike converges since 
the spike timescale is $\Delta t_{\rm sp} \sim (\sqrt{\lambda}M_\text{P})^{-1} \sim 10M_\text{P}^{-1}$.
Numerically it is checked that the Bogoliubov coefficients approach to constant values 
within this time interval, signaling the adiabaticity of the produced particles.
This is probably because we treat the very high momentum modes.
}
The result is shown in Fig.~\ref{fig:fzeta_rhozeta}.
The left and right panels show the occupation number 
$f_{\zeta_c}$ and the energy density spectrum $\dd\rho_{\zeta_c}/\dd\ln k \sim k^3\omega_kf_{\zeta_c}$.
For the total energy density, we numerically find $\rho_\zeta \simeq C_\zeta \lambda^2M_\text{P}^4/\xi^4$ 
with $C_\zeta \simeq 7\times10^{-4}$.
This value of $C_\zeta$ may come from the following: 
the factor $\bar{m}_\chi \simeq 0.5 \sqrt{\lambda}M_P/\xi$
in Eqs.~(\ref{eq:pp_f_summary}) and (\ref{eq:pp_drho_summary}),
numerical suppression in Eq.~(\ref{eq:pp_rho_summary}) and the ${\mathcal O}(1)$ factor 
mentioned in footnote~\ref{fn:f_drho}.
Also, we numerically checked that this constant does not depend on other parameters significantly 
as long as $\xi \gg 1$.
One sees that, though particles with momenta as high as $\sim \sqrt{\lambda}M_\text{P}$ are mainly produced,
their energy density is far below that of the inflaton oscillation 
$\rho_\phi\sim \lambda M_\text{P}^2 \Phi^2/\xi^2 \sim \lambda M_\text{P}^4/\xi^2$
for $\Phi \sim M_\text{P}$.
In fact, the violation of adiabaticity for this mode is quite small:
one sees that the violation is 
\begin{align}
\frac{\dot{\omega}_k}{\omega_k^2}
&\sim \frac{m_{\zeta_c}\dot{m}_{\zeta_c}}{(k^2 + m_{\zeta_c}^2)^{3/2}}
\sim \frac{(\lambda/\xi^2)M_\text{P}^2/\Delta t_{\rm sp}}{\lambda^{3/2}M_\text{P}^3}
\sim \frac{1}{\xi^2}
\ll 1,
\end{align}
even around the spike.

Here we briefly comment on Ref.~\cite{Tsujikawa:1999me}.
They analyzed particle production in the same setup in the Jordan frame,
and concluded that violent inflaton particle production occurs.
Our results do not agree with theirs, and the reason is that 
they did not take into account the mixing between the inflaton and the scalar component of the metric.
In App.~\ref{app:frame_gauge}, we have also analyzed the present system in the Jordan frame,
and confirmed that our result is independent of the flames 
if one properly takes the mixing into account.

%%%%%%%%%%%%%%%%
%%%%%%%%%%%%%%%%

%%%%%%%%%%%%%%%%
%%%%%%%%%%%%%%%%

%%%%%%%%%%%%%%%%%%%%%%%%%%%%%%%%%%%%%%%%%%%%%%%%%%
\subsubsection*{Minimal scalar}
%%%%%%%%%%%%%%%%%%%%%%%%%%%%%%%%%%%%%%%%%%%%%%%%%%

Next we consider a light scalar field $\chi_J$ whose kinetic term is minimal in the Jordan frame:
\begin{align}
	S_\chi = -\frac{1}{2}\int \dd^4x \sqrt{-g_J}\;g^{\mu\nu}_{J}\partial_{\mu}\chi_J\partial_{\nu}\chi_J,
\end{align}
where we neglect the bare mass for $\chi_J$. After moving to the Einstein frame, 
it is rewritten as
\begin{align}
	S_\chi 
	&= \int \dd\tau \dd^3x\; \frac{a^2}{2\Omega^2}\left[ \left(\chi_J'\right)^2 - \left(\partial_i \chi_J\right)^2\right].
\end{align}
In order to make the kinetic term to be canonical in the Einstein frame,
we rescale $\chi_J$ as
\begin{align}
	\chi \equiv \frac{a}{\Omega}\chi_J.
\end{align}
Then, $\chi$ obtains a time-dependent mass term as (see App.~\ref{app:frame})
\begin{align}
	m_\chi^2 
	&= 
	\frac{\Omega''}{\Omega} - \frac{2\Omega'^2}{\Omega^2}
	+ 2{\mathcal H}\frac{\Omega'}{\Omega}
	- {\mathcal H}' - {\mathcal H}^2.
	\label{eq:mass_min}
\end{align}
Thus, $\chi$ effectively couples to $\ddot{\Omega}$.
As shown in Eq.~(\ref{ddotOmega}), $\ddot\Omega$ shows a spike-like feature,
hence significant particle production is expected.

\begin{figure}[t]
\begin{center}
\includegraphics[scale=0.9]{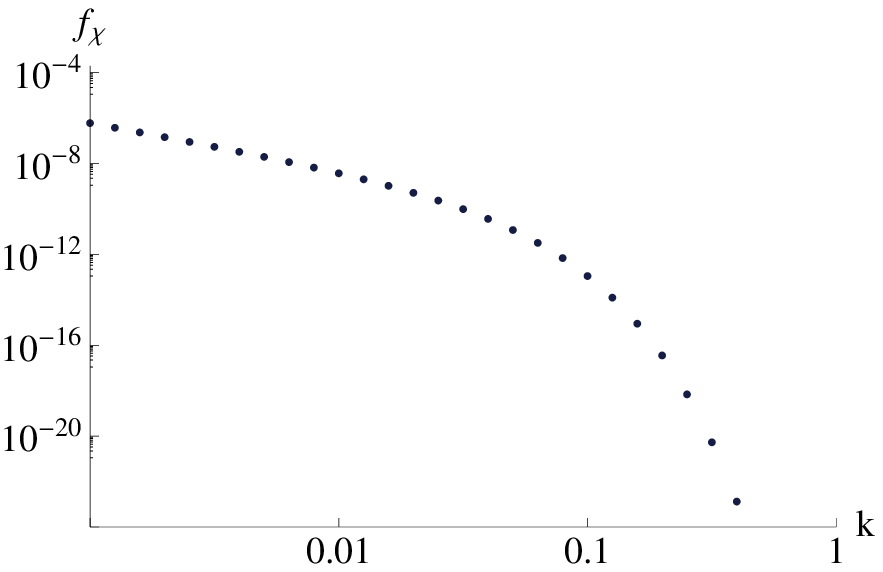}
\includegraphics[scale=0.9]{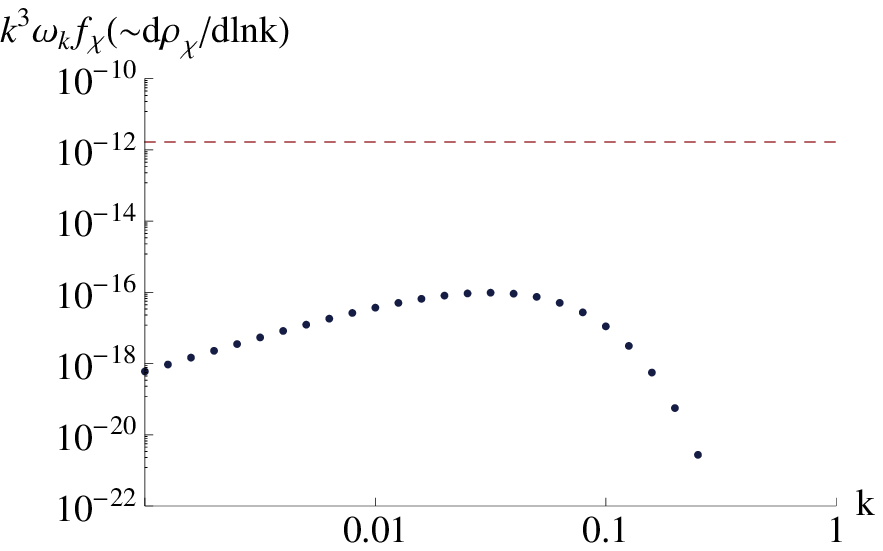}
\caption{ \small
The occupation number $f_\chi$ (left) and the energy density spectrum 
$k^3 \omega_k f_\chi$ (right, \textcolor{dark_blue}{blue}) of $\chi$ produced by the first zero-crossing of $\phi$.
The parameters are $\lambda = 0.01$ and $\xi = 10^4$,
and the initial conditions are $\phi_{\rm ini} = M_\text{P}$ and $\dot{\phi}_{\rm ini} = 0$.
In the right panel, the inflaton background energy density at the first zero-crossing is shown as
\textcolor{dark_red}{the red line}. We take $M_\text{P} = 1$ in this plot.
}
\label{fig:fchi_rhochi}
\end{center}
\end{figure}

To confirm this, 
we calculated the produced amount of $\chi$ during the first zero-crossing.
We numerically solved Eq.~\eqref{eq:EOM_alphabeta} with Eq.~\eqref{eq:mass_min}
neglecting the overall scale factor in the expression of $m_\chi^2$,
and evaluated the Bogoliubov coefficients well after the spike as in Sec.~\ref{sec:real}.
We took $\lambda = 0.01$ and $\xi = 10^4$,
and also the initial condition of the inflaton as $\phi = M_\text{P}$ with vanishing velocity.
The result is shown in Fig.~\ref{fig:fchi_rhochi}.
The left panel is the occupation number $f_\chi$ while the right panel is 
the energy density spectrum $\dd\rho_\chi / \dd\ln k \sim k^3\omega_k f_\chi$ 
\textcolor{dark_blue}{(blue)} and the inflaton energy density 
at its first zero-crossing \textcolor{dark_red}{(red)}.
As an order estimation, the occupation number $f_\chi$ coincides well 
with our analytical formula~\eqref{eq:pp_f_summary} with
\begin{align}
	\bar{m}_\chi \sim \sqrt{\frac{\lambda}{\xi}}\Phi \sim  \sqrt{\frac{\lambda}{\xi}}M_\text{P},
	\label{eq:mchibar_msp}
\end{align}
as deduced from Eq.~(\ref{ddotOmega}).
The energy density of $\chi$ after the first zero-crossing of the inflaton is thus given by
\begin{align}
	\rho_\chi \sim \frac{\lambda^2}{\xi^2}\Phi^4 \sim \frac{\lambda^2}{\xi^2}M_\text{P}^4.
\end{align}
In the last similarity of these estimates we have substituted $\Phi\sim M_\text{P}$ just after inflation.
Numerically we find $\rho_\chi \simeq C_\chi \lambda^2M_\text{P}^4/\xi^2$ 
with $C_\chi \simeq 9 \times 10^{-6}$.
This value of $C_\chi$ may come from the following: 
the factor $\bar{m}_\chi \simeq 0.2 \sqrt{\lambda/\xi}M_P$
in Eqs.~(\ref{eq:pp_f_summary}) and (\ref{eq:pp_drho_summary}),
numerical suppression in Eq.~(\ref{eq:pp_rho_summary}) and the ${\mathcal O}(1)$ factor 
mentioned in footnote~\ref{fn:f_drho}.
Also, we numerically checked that this constant does not depend on other parameters significantly 
as long as $\xi \gg 1$.
Thus $\rho_\chi$ is smaller than the inflaton energy density
$\rho_\phi \sim \lambda M_\text{P}^4/\xi^2$ by a factor of $\lambda$.
The violation of the adiabaticity for $\chi$ particles with momentum $\sim \sqrt{\lambda}M_\text{P}$
is estimated as
\begin{align}
\frac{\dot{\omega}_k}{\omega_k^2}
&\sim \frac{m_\chi\dot{m}_\chi}{(k^2 + m_\chi^2)^{3/2}}
\sim \frac{(\lambda/\xi)M_\text{P}^2/\Delta t_{\rm sp}}{\lambda^{3/2}M_\text{P}^3}
\sim \frac{1}{\xi}
\ll 1,
\end{align}
around the first spike.
However, one sees that $\chi$ particles can carry away
a nonnegligible fraction of the inflaton energy density for a sizable value of $\lambda$.

Here we comment on a nonminimal coupling between $\chi$ and $R_J$. If there is such a coupling,
\begin{align}
	S_\chi = -\frac{1}{2}\int \dd^4x \sqrt{-g_J}\left[g^{\mu\nu}_{J}\partial_{\mu}\chi_J\partial_{\nu}\chi_J
	+ \frac{\xi_\chi}{2} R_J \chi^2\right],
\end{align}
the light field $\chi$ obtains a mass term of $\mathcal{O}(\xi_\chi)$ larger than Eq.~\eqref{eq:mass_min}
(see Eq.~\eqref{eq:R_Weyl}), and hence the production of $\chi$ is enhanced by that factor.
Actually, in the global U(1) case we discuss in the next section, 
the production of the U(1) partner of the inflaton may be viewed as taking $\xi_\chi = \xi$
in the present case (up to contributions from the potential that are negligible).

%%%%%%%%%%%%%%%%
%%%%%%%%%%%%%%%%

%%%%%%%%%%%%%%%%%%%%%%%%%%%%%%%%%%%%%%%%%%%%%%%%%%
\subsection{Complex scalar inflaton -- global U(1) case}
\label{sec:global_U1}
%%%%%%%%%%%%%%%%%%%%%%%%%%%%%%%%%%%%%%%%%%%%%%%%%%

Next we consider the case where the inflaton is a part of a complex scalar field $\phi$ with a global U(1) charge.
We parametrize $\phi$ as $\phi = \phi_re^{i\theta}/\sqrt{2}$, where
the radial component $\phi_r$ is identified as the inflaton.
The production of the inflaton particle (or $\zeta$) is the same as in Sec.~\ref{sec:real},
and hence we concentrate on particle production of $\theta$ in this subsection.
We will see that the production of $\theta$ is so violent
that we should take into account the back-reaction of the produced $\theta$ particles.

We consider the following action:
\begin{align}
S
&= \int \dd^4x \sqrt{-g_J}
\left[ \left( \frac{M_\text{P}^2}{2} + \xi|\phi_J|^2 \right)R_J
- g^{\mu \nu}_J (\partial_\mu \phi_J)^\dagger (\partial_\nu \phi_J) - V_J
\right]
\nonumber  \\
&= \int \dd^4x \sqrt{-g_J}
\left[ \left( \frac{M_\text{P}^2}{2} + \frac{\xi}{2}\phi_{Jr}^2\right)R_J
	- \frac{1}{2}g^{\mu\nu}_{J}\left(\partial_\mu \phi_{Jr} \partial_\nu \phi_{Jr} 
	+ \phi_{Jr}^2\partial_\mu \theta_J \partial_\nu \theta_J\right)
- \frac{\lambda}{4}\phi_{Jr}^4
\right].
\label{eq:pp_S_global}
\end{align}
We move to the Einstein frame 
as we have done in Sec.~\ref{sec_background}.
We take the conformal factor as
\begin{align}
	\Omega^2 = 1 + \frac{\xi \phi_{Jr}^2}{M_\text{P}^2}.
\end{align}
The action is rewritten as
\begin{align}
	S =& \int \dd^4x \sqrt{-g}
	\left[ 
	\frac{M_\text{P}^2}{2} R
	-\frac{1}{2}g^{\mu\nu}\partial_\mu \phi_r \partial_\nu \phi_r - V(\phi_r)
	- \frac{\phi_{Jr}^2}{2\Omega^2}g^{\mu\nu}\partial_\mu \theta_J \partial_\nu \theta_J
	\right],
\end{align}
where the definitions of $\phi_r$ and $V(\phi_r)$ are the same as 
Eqs.~\eqref{eq:B_VE} and~\eqref{eq:mapping}.
After canonically normalizing $\theta_J$ as 
\begin{align}
	\theta \equiv \frac{a\phi_{Jr}}{\Omega}\theta_J,
\end{align}
the mass term of $\theta$ is given by (see App.~\ref{app:frame}, 
and use Eq.~(\ref{eq:app:meff_V4}))
\begin{align}
	m_\theta^2 
	&= 
	- \frac{(a\phi_{Jr}/\Omega)''}{a\phi_{Jr}/\Omega}
	=
	a^2\frac{m_{J{\rm eff}}^2}{\Omega^2}
	- \frac{a''}{a} - \frac{(1/\Omega)''}{(1/\Omega)} - \frac{2a'(1/\Omega)'}{a(1/\Omega)}.
	\label{eq:mass_phase}
\end{align}
Thus, the phase component $\theta$ couples to $m_{J{\rm eff}}^2$.
As found in Eq.~(\ref{mJeff}), $m_{J{\rm eff}}^2$ exhibits a strong spike-like feature
and hence $\theta$ is efficiently excited.\footnote{
	Precisely, $\theta$ is ill-defined at $\phi_r=0$, but it does not matter.
	Actually we obtain the same result for the production of $\phi_I$
	if we decompose $\phi$ as $\phi=(\phi_R+i\phi_I)\sqrt{2}$ and
	identify $\phi_R$ as the inflaton.
	Take the limit $g \rightarrow 0$ in Eq.~\eqref{eq:goldstone_gauge} in App.~\ref{app:frame_gauge}.
}
The $\Omega''$ term in (\ref{eq:mass_phase}) also has a spike, but it is weaker than 
the $m_{J{\rm eff}}^2$ term.

\begin{figure}[t]
\begin{center}
\includegraphics[scale=0.9]{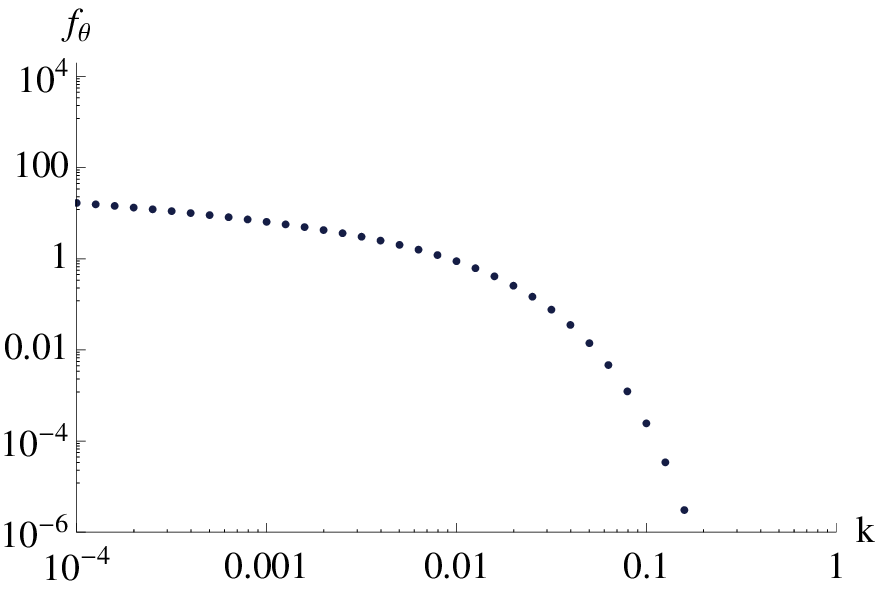}
\includegraphics[scale=0.9]{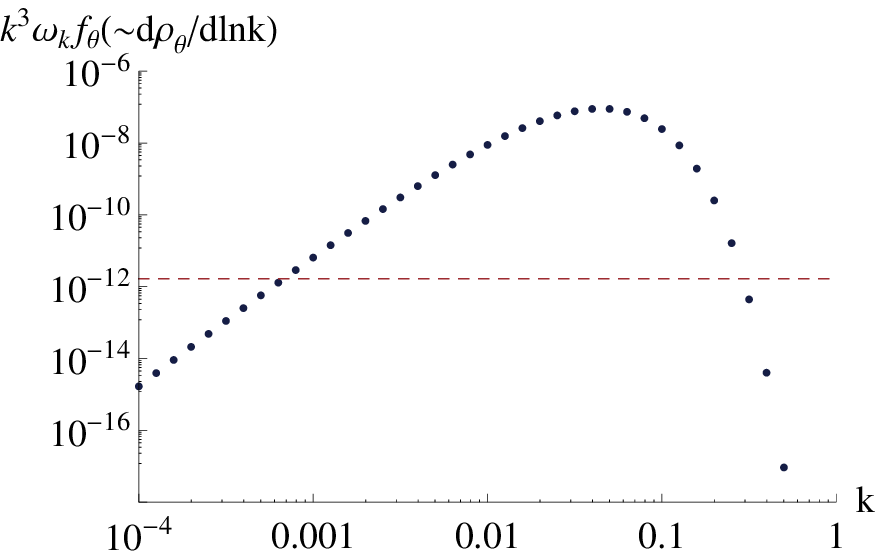}
\caption{ \small
The occupation number$f_\theta$ (left) and the energy density spectrum 
$k^3 \omega_k f_\theta$ (right, \textcolor{dark_blue}{blue}) of $\theta$
after the first zero-crossing of $\phi_r$.
The parameters and initial conditions as well as \textcolor{dark_red}{the red line} 
in the right panel are the same as in Fig.~\ref{fig:fchi_rhochi}.
We take $M_\text{P} = 1$ in this plot.
}
\label{fig:ftheta_rhotheta}
\end{center}
\end{figure}

Now we calculate the produced amount of $\theta$.
In Fig.~\ref{fig:ftheta_rhotheta}, we have numerically solved Eq.~\eqref{eq:EOM_alphabeta} to obtain
the occupation number of $\theta$ by substituting Eq.~\eqref{eq:mass_phase}
neglecting the overall scale factor in the expression of $m_\theta^2$,
and evaluated the Bogoliubov coefficients well after the first spike as in Sec.~\ref{sec:real}.
We took the same parameter values as in previous subsections.
In the left and right panels, we plot the occupation number $f_{\theta}$
and the energy spectrum $\dd\rho_{\theta}/\dd\ln k \sim k^3 \omega_k f_\theta$ 
just after the first zero-crossing of the inflaton, respectively.
Again as an order estimation, it qualitatively coincides with our analytical formula~\eqref{eq:pp_f_summary} with\footnote{
	In the present case, we have $\bar{m}_\theta \sim m_{\rm sp}$, 
	and our formula (\ref{eq:pp_f_summary}) and (\ref{eq:pp_drho_summary}) can be marginally applied.
	In fact, the $k$-dependence of $f_\theta$ now deviates from $k^{-2}$ in the left panel of Fig.~\ref{fig:ftheta_rhotheta}.
	But still the distribution is peaked around $k \sim m_{\rm sp}$ and
	it does not change the main discussion in the text.
} 
\begin{align}
	\bar{m}_\theta \sim \frac{\sqrt{\lambda} \xi \Phi_{Jr}^2}{M_\text{P}} 
	\sim \sqrt{\lambda}M_\text{P},
\end{align}
as deduced from Eq.~(\ref{mJeff}).
The energy density of $\theta$ after the first passage of $\phi=0$ is then estimated as
\begin{align}
	\rho_\theta \sim \frac{\lambda^2 \xi^4\Phi_{Jr}^8}{M_\text{P}^4} 
	\sim \lambda^2M_\text{P}^4,
\end{align}
where in the second similarity we have substituted $\Phi_{Jr} \sim M_\text{P}/\sqrt{\xi}$ just after inflation.
Numerically we find $\rho_\theta \simeq C_\theta \lambda^2M_\text{P}^4$ 
with $C_\theta \simeq 7 \times 10^{-5}$.
This value of $C_\theta$ may come from the following: 
the factor $\bar{m}_\chi \simeq 0.4 \sqrt{\lambda/\xi}M_P$
in Eqs.~(\ref{eq:pp_f_summary}) and (\ref{eq:pp_drho_summary}),
numerical suppression in Eq.~(\ref{eq:pp_rho_summary}) and the ${\mathcal O}(1)$ factor 
mentioned in footnote~\ref{fn:f_drho}.
Also, we numerically checked that this constant does not depend on other parameters significantly 
as long as $\xi \gg 1$.
The violation of the adiabaticity for $\theta$ particles with momentum $\sim \sqrt{\lambda}M_\text{P}$
is estimated as
\begin{align}
\frac{\dot{\omega}_k}{\omega_k^2}
&\sim \frac{m_\theta\dot{m}_\theta}{(k^2 + m_\theta^2)^{3/2}}
\sim \frac{\lambda M_\text{P}^2/\Delta t_{\rm sp}}{\lambda^{3/2}M_\text{P}^3}
\sim 1,
\end{align}
around the first spike. 
This means that the adiabaticity is at most marginally broken for these modes,
but nevertheless this leads to significant consequences as we see below.

Let us compare $\rho_{\theta}$ with the energy density of the inflaton $\rho_{\phi_r}$.
Taking account of the numerical factor from numerical calculation, the ratio is given by
\begin{align}
	\frac{\rho_{\theta}}{\rho_{\phi_r}} \simeq 4\times 10^{-3} \xi^2 \lambda.
\end{align}
We have to take into account the back-reaction of the produced $\theta$ to the background evolution
if the condition $\rho_\theta/\rho_{\phi_r} \gtrsim 1$ is satisfied.
We find that this condition reduces to\footnote{
	This conclusion may be different from Ref.~\cite{DeCross:2015uza}, 
	although the calculation itself is consistent in the overlapping region.
	The reason is probably that they concentrate on whether the adiabaticity can be violently broken or not,
	while we point out that the energy density of the produced particles can be sizable even if the adiabaticity is only marginally broken.
	This is mainly because the typical momentum of the produced particles $(\sim \sqrt{\lambda}M_\text{P})$
	is extremely high.
	(Note added: In the arXiv version 2 of Ref.~\cite{DeCross:2015uza}, the conclusion is consistent with ours.)
}
\begin{align}
	\xi^2 \lambda \gtrsim 3 \times 10^2.
	\label{eq:cond_back}
\end{align}
Notice that the parameters must satisfy
\begin{align}
	\xi \simeq 5\times 10^{4}\sqrt{\lambda},
\end{align}
to reproduce the observed density perturbation at cosmological scales. 
Then, Eq.~\eqref{eq:cond_back} is rewritten as
\begin{align}
	\xi \gtrsim 9\times 10^2 ~~\mathrm{or} ~~ \lambda \gtrsim 3\times10^{-4}.
\end{align}
We can see that it is actually the case for the parameters of Fig.~\ref{fig:ftheta_rhotheta}
(compare \textcolor{dark_blue}{the blue points} and \textcolor{dark_red}{the red line} in the right panel).
Note that we have neglected the back-reaction in Fig.~\ref{fig:ftheta_rhotheta}.
Once the back-reaction becomes important, one might expect that 
an $\mathcal{O}(1)$ fraction of the inflaton energy density is transferred to particles, 
and the inflaton coherent oscillation is broken,
although a thorough study on this respect is beyond the scope of this paper.

%%%%%%%%%%%%%%%%
%%%%%%%%%%%%%%%%

%%%%%%%%%%%%%%%%%%%%%%%%%%%%%%%%%%%%%%%%%%%%%%%%%%
\subsection{Complex scalar inflaton -- gauged U(1) case}
\label{sec:gauged_U1}
%%%%%%%%%%%%%%%%%%%%%%%%%%%%%%%%%%%%%%%%%%%%%%%%%%

Now we discuss the case where the inflaton has a gauged U(1) charge.
We consider the following action:
\begin{align}
S
&=
\int \dd^4x \sqrt{-g_J}
\left[ 
\left( \frac{M_\text{P}^2}{2} + \xi |\phi_J|^2 \right) R_J
- \frac{1}{4}F_{\mu \nu}F^{\mu \nu}
- (D_\mu \phi_J)^\dagger (D^\mu \phi_J)
- V_J(|\phi_J|^2)
\right],
\label{eq:pp_gauge_S}
\end{align}
where $F_{\mu\nu}$ is the field strength, $D_\mu = \partial_\mu - igA_\mu$,
$A_\mu$ is the gauge field and $g$ is the gauge coupling.
We proceed with the unitary gauge $\phi_J = \phi_{Jr} /\sqrt{2}$
where $\phi_{Jr}$ is the radial component of $\phi_J$.\footnote{
	This gauge condition is ill-defined at $\phi_r = 0$, but
	we explicitly show in App.~\ref{app:gauge} that the result is the same in the Coulomb gauge.
	Note that the Coulomb gauge is well-defined even at $\phi_r = 0$,
	and hence the ill-definition at $\phi_r = 0$ does not spoil the discussion here.
} The production of the inflaton particle (or $\zeta$) is the same as in Sec.~\ref{sec:real},
and hence we focus on the production of the gauge boson in this subsection.

By using the conformal time $\tau$,
we obtain the action for the gauge boson in the Einstein frame as
\begin{align}
	S_A = \int \dd\tau \dd^3x \left[ 
	-\frac{1}{4}\eta^{\mu\rho}\eta^{\nu\sigma}F_{\mu\nu}F_{\rho\sigma}
	-\frac{m_A^2}{2}\eta^{\mu\nu}A_{\mu}A_{\nu}
	\right],
	\label{eq:action_gauge}
\end{align}
where $\eta_{\mu\nu} = \mathrm{diag}(-1, +1, +1, +1)$ and we have defined
\begin{align}
	m_A^2 \equiv \frac{a^2g^2\phi_{Jr}^2}{\Omega^2},
	~~~
	\Omega^2 = 1 + \frac{\xi \phi_{Jr}^2}{M_\text{P}^2}.
	\label{mA_Ein}
\end{align}
The relation between $\phi_{Jr}$ and the inflaton in the Einstein frame $\phi_r$ is 
the same as Eq.~\eqref{eq:mapping}.
Notice that Eq.~\eqref{eq:action_gauge} contains an unphysical degree of freedom $A_\tau$.
Thus, we should first eliminate it to obtain the physical action for the transverse and the longitudinal modes.
During that procedure, we will make it clear that the mass terms of the transverse and the longitudinal
modes are different in general if the symmetry breaking field (the inflaton in our case) is time-dependent.
After obtaining the mass term, we discuss the production of the gauge bosons, especially the longitudinal mode.

%%%%%%%%%%%%%%%%%%%%%%%%%%%%%%%%%%%%%%%%%%%%%%%%%%
\subsubsection*{Action for physical degrees of freedom}
%%%%%%%%%%%%%%%%%%%%%%%%%%%%%%%%%%%%%%%%%%%%%%%%%%

The action~\eqref{eq:action_gauge} does not contain the time derivative of $A_\tau$, and hence
it is not a dynamical degree of freedom. We have to eliminate it to obtain the action
for the transverse and the longitudinal modes. By noting that
\begin{align}
	- \frac{1}{4} \eta^{\mu\rho}\eta^{\nu\sigma}F_{\mu\nu}F_{\rho\sigma}
	=
	\frac{1}{2}\left(\partial_i A_\tau\right)^2 - A_i'\partial_i A_\tau
	+ \frac{1}{2}\left\lvert \vec{A}'\right\rvert^2 - \frac{1}{2}\left\lvert \vec{\nabla} \times \vec{A}\right\rvert^2,
\end{align}
where the prime denotes the derivative with respect to $\tau$, we obtain 
\begin{align}
	S_A = \frac{1}{2}\int \frac{\dd\tau \dd^3k}{\left(2\pi\right)^3}
	\Bigg[
	\left(k^2 + m_A^2\right)\left\lvert A_\tau + \frac{i\vec{k}\cdot\vec{A}'}{k^2 + m_A^2}\right\rvert^2
	~~~~~~~~~~~~~~~~~~~~~~~~~~~
	\nonumber \\[.4em]
	+ \left\lvert \vec{A}'\right\rvert^2 - \left\lvert \vec{k}\times \vec{A}\right\rvert^2
	- \frac{\left\lvert \vec{k}\cdot\vec{A}'\right\rvert^2}{k^2 + m_A^2} - m_A^2 \left\lvert \vec{A}\right\rvert^2
	\Bigg],
\end{align}
where we have moved on to the Fourier space.
After eliminating $A_\tau$, we can reduce the action to
\begin{align}
	S_A = \frac{1}{2}\int \frac{\dd\tau \dd^3k}{\left(2\pi\right)^3}
	\left[	\left\lvert \vec{A}^\prime\right\rvert^2 - \left\lvert \vec{k}\times \vec{A}\right\rvert^2
	- \frac{\left\lvert \vec{k}\cdot\vec{A}'\right\rvert^2}{k^2 + m_A^2} - m_A^2 \left\lvert \vec{A}\right\rvert^2
	\right].
\end{align}
Now we decompose the spatial part of the gauge boson into the transverse mode $\vec{A}_T$
and the longitudinal mode $\tilde{A}_L$ as
\begin{align}
	\vec{A} = \vec{A}_T + \frac{\vec{k}}{k}\tilde{A}_L, ~~~ \vec{k}\cdot\vec{A}_T = 0.
\end{align}
Then, the action becomes
\begin{align}
	S_A = S_{A_T} + S_{A_L},
\end{align}
where
\begin{align}
	S_{A_T} &= \frac{1}{2}\int \frac{\dd\tau \dd^3k}{\left(2\pi\right)^3}
	\left[\left\lvert\vec{A}_T'\right\rvert^2 
	- \left(k^2 + m_A^2\right)\left\lvert \vec{A}_T\right\rvert^2
	\right],
	\label{S_AT}
\end{align}
and
\begin{align}
	S_{A_L} 
	&= \frac{1}{2}\int \frac{\dd\tau \dd^3k}{\left(2\pi\right)^3}
	\left[
	\frac{m_A^2}{k^2 + m_A^2}\left\lvert \tilde{A}_L'\right\rvert^2
	- m_A^2 \left\lvert \tilde{A}_L\right\rvert^2
	\right] \nonumber \\[.4em]
	&= \frac{1}{2}\int \frac{\dd\tau \dd^3k}{\left(2\pi\right)^3}
	\left[
	\left\lvert A_L'\right\rvert^2
	-\left(k^2 + m_A^2 - \frac{k^2}{k^2 + m_A^2}\left(\frac{m_A''}{m_A}
	-\frac{3m_A'^2}{k^2 + m_A^2}\right)\right)
	\left\lvert A_L\right\rvert^2
	\right].
	\label{S_AL}
\end{align}
Here we defined the canonically normalized longitudinal mode as
\begin{align}
	A_L \equiv \frac{m_A}{\sqrt{k^2 + m_A^2}}\tilde{A}_L.
\end{align}
Thus, the mass terms for the transverse mode and the longitudinal mode
are different if $m_A$ (or the symmetry breaking field) is time-dependent.
Note that this is a generic conclusion derived from the action~\eqref{eq:action_gauge}, 
irrespective of the detailed form of $m_A$.
The effect of this mass splitting is studied in Ref.~\cite{Lozanov:2016pac}
in the context of preheating in inflation without the nonminimal coupling.\footnote{
	The same effect leads to a different amount of production 
	of the longitudinal and transverse modes 
	from inflationary quantum fluctuations~\cite{Graham:2015rva}.
} In our case, the mass term of $A_L$ contains a term proportional to $\phi_{Jr}''/\phi_{Jr}$,
or equivalently $m_{J{\rm eff}}^2$,
and hence $A_L$ is violently excited by its spike-like feature.\footnote{
	This mass splitting was overlooked in Refs.~\cite{Bezrukov:2008ut,GarciaBellido:2008ab},
	and hence the production of the longitudinal gauge boson was underestimated.
} Below we analyze the production of the transverse and longitudinal mode.

%%%%%%%%%%%%%%%%%%%%%%%%%%%%%%%%%%%%%%%%%%%%%%%%%%
\subsubsection*{Production of transverse mode}
%%%%%%%%%%%%%%%%%%%%%%%%%%%%%%%%%%%%%%%%%%%%%%%%%%

As seen from the action (\ref{S_AT}), the transverse mode $A_T$ has just a mass of $m_A^2$ defined in (\ref{mA_Ein}).
Since the time dependence of $\phi_{Jr}$ itself does not have a strong spike, the production of transverse mode is similar to the
standard analysis of broad resonance in the Einstein frame~\cite{Kofman:1997yn}.

Let us derive the typical momentum of produced gauge boson. 
The transverse gauge boson mass is written as
\begin{align}
	m_{A_T}^2 = \frac{a^2 g^2 \phi_{Jr}^2}{\Omega^2} \sim \frac{g^2 M_\text{P}}{\xi}|\phi|.
\end{align}
We assume $g^2\Phi^2/m_{\rm osc}^2 \gg 1$ so that the preheating happens in the broad regime~\cite{Kofman:1997yn}.
The background evolution becomes non-adiabatic for gauge bosons,
if the condition $|\dot\omega_k/\omega_k^2| \ll 1$ is violated, where $\omega_k^2=k^2+m_{A_T}^2$.
The non-adiabatic change leads to copious particle production
with $k<k_*$~\cite{Bezrukov:2008ut,GarciaBellido:2008ab}:
\begin{align}
	k_* \simeq \left( \frac{g^2M_\text{P} m_{\rm osc} \Phi}{\xi} \right)^{1/3}.
\end{align}
Therefore the energy density of the transverse gauge boson after the first zero-crossing of the inflaton is estimated as
\begin{align}
	\rho_{A_T} \sim g\Phi_J k_*^3 \sim \frac{g^3 \sqrt{\lambda} M_\text{P} \Phi_J^3}{\xi} 
	\sim \frac{g^3 \sqrt{\lambda} M_\text{P}^4}{\xi^{5/2}}.
\end{align}
Here we have used 
the fact that the produced gauge bosons are non-relativistic except for the small time interval around $\phi_J \simeq 0$.
In the last similarity we have used $\Phi_J \sim M_\text{P}/\sqrt{\xi}$ just after inflation.
Below we see that the longitudinal gauge boson after the first zero-crossing
can be much more abundant for large enough $\xi$.

%%%%%%%%%%%%%%%%%%%%%%%%%%%%%%%%%%%%%%%%%%%%%%%%%%
\subsubsection*{Production of longitudinal mode}
%%%%%%%%%%%%%%%%%%%%%%%%%%%%%%%%%%%%%%%%%%%%%%%%%%

Now we calculate the produced amount of $A_L$.
The mass term of $A_L$ is given by
\begin{align}
	m_{A_L}^2 
	&= 
	m_A^2 - \frac{k^2}{k^2 + m_A^2}\left(\frac{m_A''}{m_A}
	-\frac{3m_A'^2}{k^2 + m_A^2}\right),
	~~~ 
	m_A^2 = \frac{a^2g^2\phi_{Jr}^2}{\Omega^2}.
	\label{eq:mass_gauge}
\end{align}
Note that the spike-like feature of $m_{A_L}^2$ is determined by the $m_A''/m_A$ part:
\begin{align}
	m_{A_L}^2 
	&\sim
	\frac{k^2}{k^2 + m_A^2}\frac{m_{J{\rm eff}}^2}{\Omega^2},
	\label{eq:mass_gauge_approx}
\end{align}
The longitudinal gauge boson $A_L$ feels strong spike of $m_{J{\rm eff}}^2$
as in the global U(1) case studied in the previous subsection.
The spike-like feature of Eq.~\eqref{eq:mass_gauge} is totally the same as 
that of Eq.~\eqref{eq:mass_phase} for the small field region where $k \gg m_A$.

\begin{figure}[t]
\begin{center}
\includegraphics[scale=0.9]{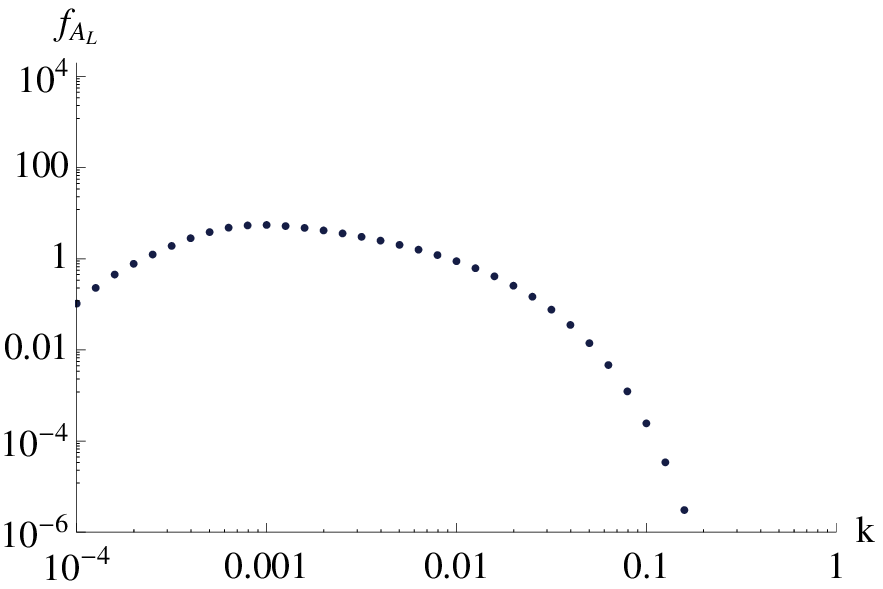}
\includegraphics[scale=0.9]{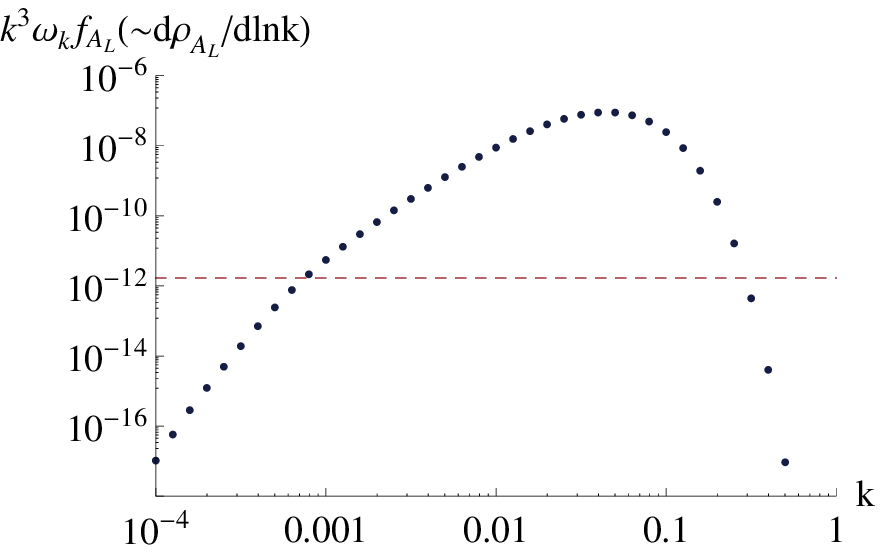}
\caption{ \small
The occupation number$f_{A_L}$ (left) and the energy density spectrum 
$k^3 \omega_k f_{A_L}$ (right, \textcolor{dark_blue}{blue}) 
of $A_L$
by the first zero-crossing of $\phi_r$.
The parameters and initial conditions as well as \textcolor{dark_red}{the red line} 
in the right panel are the same as in Fig.~\ref{fig:fchi_rhochi}.
We take $M_\text{P} = 1$ in this plot.
}
\label{fig:fAL_rhoAL}
\end{center}
\end{figure}

In Fig.~\ref{fig:fAL_rhoAL}, 
we have numerically calculated the amount of produced particles with Eq.~(\ref{eq:mass_gauge}),
approximating the overall scale factor and the comoving wavenumber $k/a$ 
to be constant during the spike.
We took the same parameter values as in previous subsections,
and estimated the Bogoliubov coefficients well after the spike.
We have taken $g = 1$, though this choice has only negligible effect on the final result
since it is cancelled in $m_A''/m_A$.
In the left and right panels, we plot the occupation number $f_{A_L}$
and the energy spectrum $\dd\rho_{A_L}/\dd\ln k \sim k^3 \omega_k f_{A_L}$ 
just after the first zero-crossing of the inflaton, respectively.
Again as an order estimation, similar to the global U(1) case, it coincides with our analytical formula~\eqref{eq:pp_f_summary} with\footnote{
	We have $\bar{m}_{A_L} \sim m_{\rm sp}$ in this case, 
	and our formula (\ref{eq:pp_f_summary}) and (\ref{eq:pp_drho_summary}) can be again marginally applied.
} 
\begin{align}
	\bar{m}_{A_L} \sim  \frac{\sqrt{\lambda} \xi \Phi_{Jr}^2}{M_\text{P}}
	\sim \sqrt{\lambda}M_\text{P}, 
\end{align}
and the energy density of $A_L$ is estimated as
\begin{align}
	\rho_{A_L} \sim \frac{\lambda^2 \xi^4 \Phi_{Jr}^8}{M_\text{P}^4} 
	\sim \lambda^2M_\text{P}^4,
\end{align}
where we have substituted $\Phi_{Jr} \sim M_\text{P}/\sqrt\xi$ just after inflation in the second similarity.
This energy density is mainly contributed from particles with momenta $\sim \sqrt{\lambda}M_\text{P}$.\footnote{
For gauge bosons with these momenta, perturbative decay just after the spike is negligible.
This is because the mass of the gauge bosons coming from the inflaton expectation value
$g^2 \phi^2 \sim g^2 M_\text{P}^2/\xi^2$, and hence the decay rate as well, are much smaller than
the spike time scale inverse just after the spike.
For perturbative decay much after the spike,
we have to take into account the back-reaction to the inflaton motion since it determines the gauge boson mass,
and therefore such study is beyond the scope of this paper.
}
Numerically we find $\rho_{A_L} \simeq C_{A_L} \lambda^2M_\text{P}^4$  
with $C_{A_L} \simeq 7 \times 10^{-5}$.
This value of $C_{A_L}$ may come from the following: 
the factor $\bar{m}_\chi \simeq 0.4 \sqrt{\lambda/\xi}M_P$
in Eqs.~(\ref{eq:pp_f_summary}) and (\ref{eq:pp_drho_summary}),
numerical suppression in Eq.~(\ref{eq:pp_rho_summary}) and the ${\mathcal O}(1)$ factor 
mentioned in footnote~\ref{fn:f_drho}.
Also, we numerically checked that this constant does not depend on other parameters significantly 
as long as $\xi \gg 1$.
This result is the same as $\rho_\theta$ in the global U(1) case,
and the discussion on the violation of the adiabaticity is also the same.
The only difference from the global U(1) case is 
in the spectral bent around $k \sim 10^{-3}M_P$ in Fig.~\ref{fig:fAL_rhoAL},
and this corresponds to the wavenumber $k \sim gM_P/\xi$
below which the factor $k^2/(k^2 + m_A^2)$ remains much below the unity
even after the inflaton enters the narrow region $\lvert \phi \rvert \ll M_P/\xi$.
The ratio of the energy densities of $A_L$ and the inflaton is again given by 
\begin{align}
	\frac{\rho_{A_L}}{\rho_{\phi_r}} \simeq 4\times 10^{-3} \xi^2 \lambda,
\end{align}
and we have to take into account the back-reaction of $A_L$
if the condition $\rho_{A_L}/\rho_\phi \gtrsim 1$ is satisfied.
%Taking account of numerical factors obtained at the reference point $\lambda = 0.01$ and $\xi = 10^4$, 
We find that this condition reduces to
\begin{align}
	\xi^2 \lambda \gtrsim 3\times10^2,
	\label{eq:cond_back_gauge}
\end{align}
is satisfied. Again, by requiring consistency with observations, 
we can rewrite Eq.~\eqref{eq:cond_back_gauge} as
\begin{align}
	\xi \gtrsim 9\times 10^2 ~~\mathrm{or} ~~ \lambda \gtrsim 3\times10^{-4}.
\end{align}
This is actually the case for the parameters of Fig.~\ref{fig:fAL_rhoAL}
(compare \textcolor{dark_blue}{the blue points} and \textcolor{dark_red}{the red line} in the right panel).
Note again that we have neglected the back-reaction in Fig.~\ref{fig:fAL_rhoAL}.
Once the back-reaction becomes important, one might expect that 
an $\mathcal{O}(1)$ fraction of the inflaton energy density is transferred to $A_L$, 
and the inflaton coherent oscillation is broken,
although a thorough study is again beyond the scope of this paper.

Here we comment on the case of the SM Higgs inflation model.
Since $\lambda\sim 0.01$ for the SM Higgs,
we naively expect that the production of the longitudinal gauge boson is so violent that the 
almost all the inflaton energy density is transferred to high momentum gauge bosons within one inflaton oscillation,
making the precise analysis thereafter quite complicated.\footnote{
This conclusion can change in the critical Higgs inflation~\cite{Hamada:2014iga,Bezrukov:2014bra}, 
where $\xi \sim {\mathcal O}(10)$.
}
To be more rigorous, we must analyze the system in the non-Abelian gauge group case,
although we expect a similar feature because the self interaction of the gauge bosons 
is safely neglected at the beginning of the preheating stage.
We leave a detailed analysis for a future work.

%%%%%%%%%%%%%%%%
%%%%%%%%%%%%%%%%

%%%%%%%%%%%%%%%%%%%%%%%%%%%%%%%%%%%%%%%%%%%%%%%%%%
\section{Conclusions and discussion}
\label{sec_conc}
\setcounter{equation}{0}
%%%%%%%%%%%%%%%%%%%%%%%%%%%%%%%%%%%%%%%%%%%%%%%%%%

In the present paper, we have analyzed the inflaton oscillating regime, 
or the preheating regime, of the inflation model with
nonminimal coupling $\xi \phi^2 R$ between the inflaton $\phi$ and the Ricci scalar $R$, 
and with quartic potential $\lambda \phi^4/4$.
We have pointed out that there are two typical mass scales (or inverse time scales)
in the dynamics of the inflaton for $M_\text{P}/\xi \ll \Phi \ll M_\text{P}$ where
$\Phi$ is the inflaton oscillation amplitude in the Einstein frame.
One is the inflaton oscillation scale,
\begin{align}
	m_\mathrm{osc} = \Delta t_\mathrm{osc}^{-1} \sim \frac{\sqrt{\lambda}M_\text{P}}{\xi},
\end{align}
and the other is a much shorter time scale which we call the ``spike'' scale,
\begin{align}
	m_\mathrm{sp} = \Delta t_\mathrm{sp}^{-1} \sim \sqrt{\lambda}\Phi,
\end{align}
which can be as high as $m_\mathrm{sp} \sim \sqrt{\lambda} M_\text{P}$ at the beginning of the 
inflaton oscillation.
This time scale $\Delta t_\mathrm{sp}$ corresponds to the time interval at which the inflaton passes through 
the region with $\lvert \phi_J \rvert \sim \lvert \phi \rvert \lesssim M_\text{P}/\xi$ 
where $\phi$ ($\phi_J$) is the inflaton in the Einstein frame (Jordan frame).
The spike scale appears in the dynamics of $\phi_J$ as a change of the kinetic term,
while it appears in the dynamics of $\phi$ as a sudden change of the shape of the potential.
We have found that several quantities such as the inflaton mass scale
in the Jordan frame $m_{J\mathrm{eff}}^2$ or the conformal factor 
blow up during this time interval $\Delta t_\mathrm{sp}$ (see Figs.~\ref{fig:phi_mJ2}
and~\ref{fig:ddOmega}).
The mass scale of this blow-up $m_\mathrm{sp}$ is extremely high, 
$m_\mathrm{sp} \sim \sqrt{\lambda} M_\text{P}$ at the beginning of the inflaton oscillation,
and this is why we call this feature as a ``spike'' in this paper.
Some light particles inevitably couple to this spike-like feature,
and hence we have studied particle production caused by it.
What we have found is the following:
\begin{itemize}

\item
In the case where the inflaton is a real scalar, 
the inflaton particle itself couples to the spike-like feature.
Inflaton particles with momentum $\sim m_{\rm sp}$ are produced,
though their energy density is far below that of the inflaton oscillation.
If one introduce an additional light scalar field, 
it may be sizably produced by the spike-like feature
of the conformal factor.

\item
In the case where the inflaton has a global U(1) charge,
the production of the U(1) partner $\theta$ of the inflaton is so efficient that 
almost all the initial inflaton energy density is converted to $\theta$
within one oscillation for $\xi \gtrsim 9\times10^{2}$ or $\lambda \gtrsim 3\times10^{-4}$.

\item
In the case where the inflaton has a gauged U(1) charge,
the longitudinal component of the gauge field plays the same role 
as $\theta$ in the global U(1) case.
\end{itemize}
Thus, the preheating dynamics of the inflation model 
with the nonminimal coupling to the gravity
can be much more violent than previously thought.

In this paper, we investigated only the very first stage of the preheating just after inflation.
In order to fully understand the reheating phenomena,
we must first take into account the back-reaction effects, which is expected to be important
when Eqs.~(\ref{eq:cond_back}) or (\ref{eq:cond_back_gauge}) are satisfied.
On top of that, we must take account of 
resonant amplification of the coupled fields and thermalization processes after that. 
A typical situation is that produced particles decay into lighter particles leading to 
efficient production of thermal plasma~\cite{Felder:1998vq}
and then particles in thermal bath scatter off the inflaton~\cite{Mukaida:2012qn,Mukaida:2012bz}.
The reheating is completed through these processes, which, however, is significantly model dependent,
and hence it is beyond the scope of this paper to study the reheating dynamics till the end.

Most of the discussion on the particle production in the gauged U(1) case holds true even for gauged SU(2) case.
Hence our study is also relevant for the Higgs inflation: the production of gauge bosons with 
extremely high momenta are unavoidable.
At this point, one may encounter a unitarity problem.
Since the momentum scale of particles produced by the spike is extremely high ($\sim \sqrt{\lambda} M_\text{P}$), 
we must be careful on the cutoff scale of the theory~\cite{Lerner:2009na,Ferrara:2010in,Bezrukov:2010jz}.
As noted in Refs.~\cite{Ferrara:2010in,Bezrukov:2010jz}, the cutoff scale depends on the inflaton field value and 
actually we can safely treat fluctuations during inflation.
During the preheating stage, however, particles with momentum higher than the cutoff scale may be excited due to the spike,
which may imply a difficulty to describe the reheating without some UV completion~\cite{Giudice:2010ka,Lerner:2010mq,Lerner:2011it}.
It might be non-trivial to correctly estimate the cutoff of the energy scale under the rapidly oscillating background,
but readers should keep in mind that the extremely high mass scale of the spike can invalidate the analysis of (p)reheating
within the original framework,
though a phenomenological consequence that the Higgs field is thermalized at very high temperature~\cite{Bezrukov:2008ut} 
might not be affected much.
We will come back to this issue in a separate publication.

Our results heavily rely on the fact that we have started from a simple action in the Jordan frame,
\textit{i.e.}~the kinetic term and potential of the inflaton 
have minimal forms except for the nonminimal coupling $\xi |\phi|^2 R$.
Instead, we could start with a \textit{minimal} kinetic term of the inflaton in the Einstein frame with 
a special form of the potential $V(|\phi|)$ suitable for inflation,
as discussed in Ref.~\cite{Lerner:2010mq}.
In such a case, the preheating is not as violent as the Jordan-frame-originated case.
Interestingly,
even if we start from the Einstein frame action,
we can have a violent particle production due to a spike-like feature, 
once \textit{nonminimal} kinetic terms are introduced.
As an example, let us consider the following action as assumed in the running kinetic inflation model~\cite{Takahashi:2010ky,Nakayama:2010kt}:
\begin{align}
	S = \int \dd^4x \sqrt{-g}\left[\frac{M_\text{P}^2}{2}R
	-g^{\mu\nu}\partial_\mu\phi^\dagger \partial_\nu \phi 
	-\frac{1}{M^2}g^{\mu\nu}\left(\partial_\mu |\phi|^2 \right)\left(\partial_\nu |\phi|^2 \right)
	- V(|\phi|)\right].
\end{align}
In this model, the potential becomes effectively flatter for $\lvert \phi \rvert \gg M$, providing
a good candidate for the inflaton potential.
Inflation ends at around when $\lvert \phi \rvert \sim \sqrt{M_\text{P} M}$, and the inflaton oscillates 
at around the origin of the potential after that. During the inflaton oscillation regime, 
the kinetic term changes for $\lvert \phi \rvert \gg M$ and $\lvert \phi \rvert \ll M$,
and hence there must be again a spike scale in this model.
Thus, we expect that the situation is similar to the case of the inflaton dynamics in the Jordan frame.
In particular, the U(1) partner of the inflaton is expected to feel the strong spike and violently produced.
It may be interesting to study further on how the spike scale 
affects the preheating dynamics in this class of models.

Finally we comment on the Starobinsky inflation model~\cite{Starobinsky:1980te},
\begin{align}
	S = \int \dd^4x \sqrt{-g_J}\frac{M_\text{P}^2}{2}\left[ R_J + \frac{R_J^2}{2M^2}\right].
\end{align}
In this case, by introducing an auxiliary field $\phi_J$, we can rewrite the action as
\begin{align}
	S = \int \dd^4x \sqrt{-g_J}\frac{M_\text{P}^2}{2}\left[f(\phi_J) + f'(\phi_J)\left(R_J - \phi_J\right)\right],
	~~ f(\phi_J) = \phi_J + \frac{\phi_J^2}{2M^2},
\end{align}
where the prime denotes the derivative with respect to $\phi_J$.
It is clear from the above action 
that the inflaton (or scalaron) $\phi_J$ has only kinetic mixing with
the scalar component of the metric through the nonminimal coupling, and hence 
there does not appear any spike scale in this model.
Thus, although the prediction for the scalar and tensor fluctuations is similar for the Starobinsky and Higgs inflation models,
the reheating dynamics can be much different between them.

%%%%%%%%%%%%%%%%%%%%%%%%%%%%%%%%%%%%%%%%%%%%%%%%%%%%%%%
\section*{Acknowledgments}
%%%%%%%%%%%%%%%%%%%%%%%%%%%%%%%%%%%%%%%%%%%%%%%%%%%%%%%

This work was supported by the Grant-in-Aid for Scientific Research on Scientific Research A (No.26247042 [KN]),
Young Scientists B (No.26800121 [KN]) and Innovative Areas (No.26104009 [KN], No.15H05888 [KN]).
This work was supported by World Premier International Research Center Initiative (WPI Initiative), MEXT, Japan. 
The work of R.J. was supported by IBS (Project Code IBS-R018-D1).
The work of Y.E., R.J. and K.M. was supported in part by JSPS Research Fellowships for Young Scientists.
The work of Y.E. was also supported in part by the Program for Leading Graduate Schools, MEXT, Japan.

%%%%%%%%%%%%%%%%%%%%%%%%%%%%%%%%%%%%%%%%%%%%%%%%%%
\appendix
%%%%%%%%%%%%%%%%%%%%%%%%%%%%%%%%%%%%%%%%%%%%%%%%%%

%%%%%%%%%%%%%%%%%%%%%%%%%%%%%%%%%%%%%%%%%%%%%%%%%%
\section{Background dynamics in the Jordan frame}
\label{app:bg}
\setcounter{equation}{0}
%%%%%%%%%%%%%%%%%%%%%%%%%%%%%%%%%%%%%%%%%%%%%%%%%%

In this appendix, we analyze the dynamics of the inflaton in the Jordan frame,
whose equations of motion are given by Eqs.~\eqref{eq:eom1_J}, \eqref{eq:eom2_J} and~\eqref{eq:eom3_J}.
We define $\Phi_J$ as the inflaton field value when it is slow-rolling, 
while as the inflaton oscillation amplitude after it starts to oscillate.
With this definition, it is convenient to consider the following three phases separately:
(Phase~0) $M_\text{P}/\sqrt{\xi} \ll \Phi_J$, (Phase~1) $M_\text{P}/\xi \ll \Phi_J \ll M_\text{P}/\sqrt{\xi}$ and 
(Phase~2) $\Phi_J \ll M_\text{P}/\xi$.
We study the dynamics of $\phi_J$ in each phase separately in the following.
We neglect the particle production in this appendix.

%%%%%%%%%%%%%%%%%%%%%%%%%%%%%%%%%%%%%%%%%%%%%%%%%%
\subsubsection*{Phase 0: $M_\text{P}/\sqrt{\xi} \ll \Phi_J$} 
%%%%%%%%%%%%%%%%%%%%%%%%%%%%%%%%%%%%%%%%%%%%%%%%%%

Let us start with the inflaton value satisfying $\xi \phi_J^2 \gg M_\text{P}^2$.
We will see that inflation occurs in this phase.
If we neglect the terms with the time derivatives, Eqs.~\eqref{eq:eom1_J}, \eqref{eq:eom2_J} and~\eqref{eq:eom3_J} 
give $H_J^2 \simeq \lambda \phi_J^2/12\xi$, while $m_{J\mathrm{eff}}^2 \simeq \lambda M_\text{P}^2/\xi^2$.
The ratio is
\begin{align}
	\frac{m_{J\mathrm{eff}}^2}{H_J^2} \sim \frac{M_\text{P}^2}{\xi \phi_J^2} \ll 1,
\end{align}
and hence the slow-roll condition is indeed satisfied. Therefore, it is a consistent slow-roll approximation to neglect
the terms with the time derivatives.
In order to see how inflation ends, we include the first order terms in the slow-roll approximation:\footnote{
	We have neglected the $H_J \dot{\phi}_J$ term in Eq.~\eqref{eq:eom1_J} because it is smaller than
	the $\xi \dot{H}_J\phi_J$ term by $\mathcal{O}(\xi)$ as we can see from Eq.~\eqref{eq:app:infsol}.
}
\begin{align}
-\xi (6\dot{H}_J + 12H_J^2 )\phi_J + \lambda \phi_J^3
&\simeq 0,
\label{eq:app:phi_tEOM_inf} \\
3M_\text{P}^2H_J^2 + \xi(3H_J^2\phi_J^2 + 6H_J \phi_J \dot{\phi}_J)
&\simeq \frac{\lambda}{4}\phi_J^4. 
\label{eq:app:Friedmann_inf}
\end{align}
We can solve them to obtain\footnote{
They satisfy Eq.~\eqref{eq:app:phi_tEOM_inf} exactly, 
and Eq.~\eqref{eq:app:Friedmann_inf} up to second order
in the slow-roll approximation.
}
\begin{align}
\dot{H}_J
&\simeq - \frac{\lambda}{18\xi^2}M_\text{P}^2, 
\;\;\;\;\;\;
\phi^2_J
\simeq \frac{12\xi}{\lambda}H_J^2 - \frac{M_\text{P}^2}{3\xi}.
\label{eq:app:infsol}
\end{align}
The first equation describes how the Hubble parameter decreases in time, 
and it leads to the decrease in $\phi_J$ through the second equation.
This solution is valid until $\phi_J$ drops to $\xi \phi_J^2 \sim M_\text{P}^2$, when Phase 1 starts.

%%%%%%%%%%%%%%%%%%%%%%%%%%%%%%%%%%%%%%%%%%%%%%%%%%
\subsubsection*{Phase 1: $M_\text{P} / \xi \ll \Phi_J \ll M_\text{P} / \sqrt{\xi}$}
%%%%%%%%%%%%%%%%%%%%%%%%%%%%%%%%%%%%%%%%%%%%%%%%%%

\begin{figure}[t]
\begin{center}
\includegraphics[scale=1]{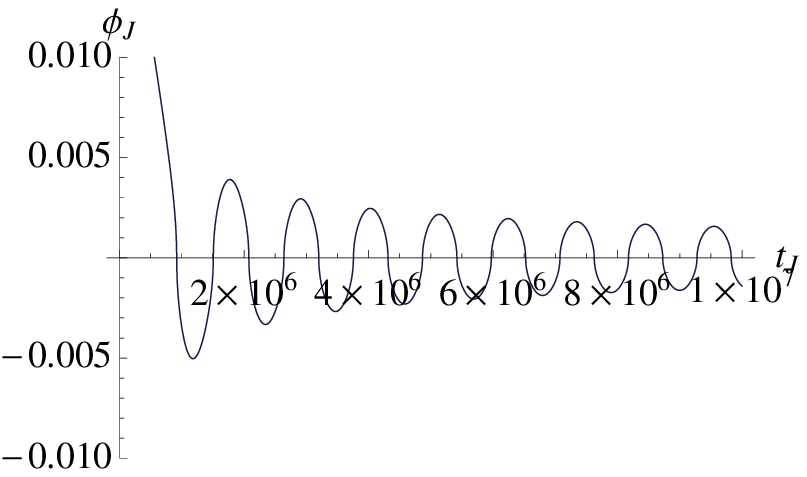}
\includegraphics[scale=1]{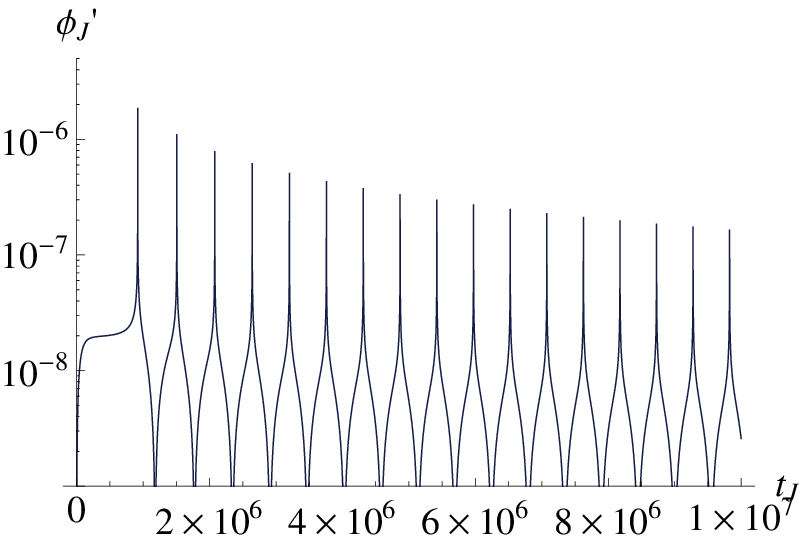}
\caption{ \small
Plot of
$\phi_J$ (left) and $|\dot{\phi}_J|$ (right) as functions of $t_J$ at the beginning of Phase 1.
Parameters are taken to be $\lambda = 0.01$ and $\xi = 10^4$,
and the initial conditions are $\phi_{J{\rm ini}} = 2M_\text{P}/\sqrt{\xi}$ and $\dot{\phi}_{J{\rm ini}} = 0$.
We take $M_\text{P} = 1$ in this plot.
}
\label{fig:phiJ_dphiJ_tJ}
\end{center}
\end{figure}

\begin{figure}[t]
\begin{center}
\includegraphics[scale=0.9]{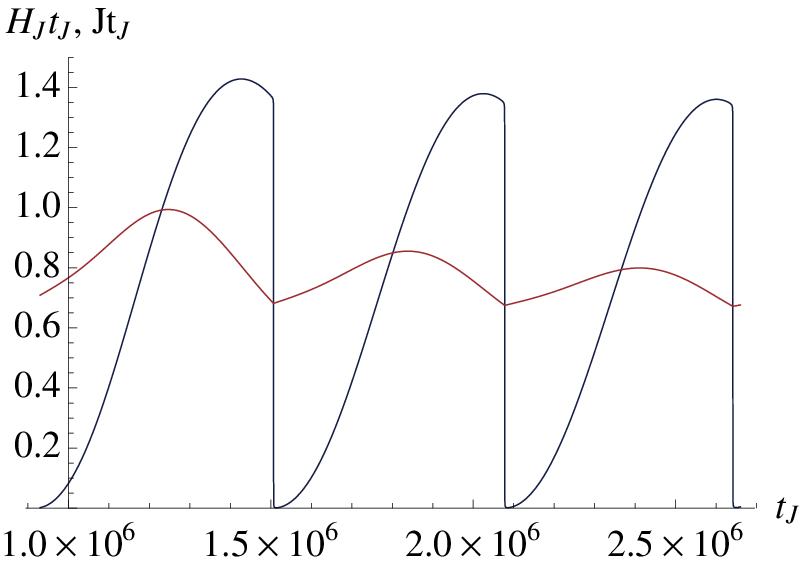}
\includegraphics[scale=0.9]{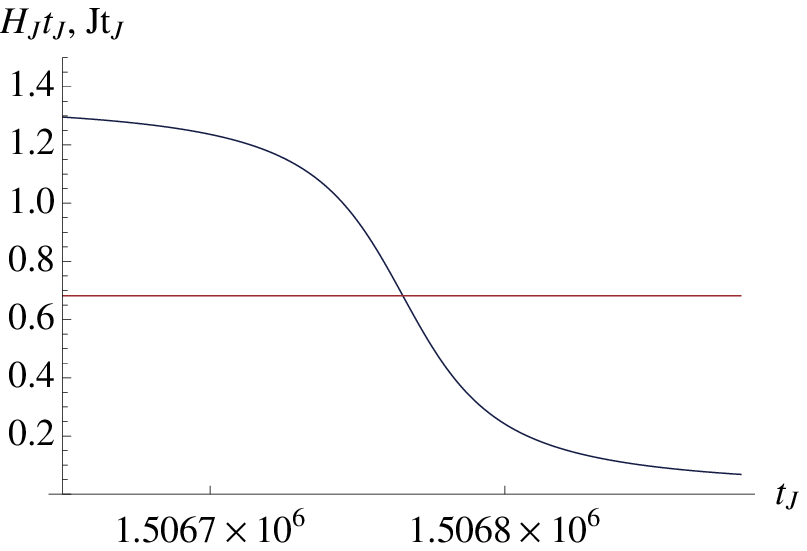}
\caption{ \small
$H_Jt_J$ (\textcolor{dark_blue}{blue}) and $Jt_J$ (\textcolor{dark_red}{red}) at the beginning of Phase 1.
Right panel is the magnification of the left panel.
The parameters and initial conditions are the same as those in Fig.~\ref{fig:phiJ_dphiJ_tJ}.
We take $M_\text{P} = 1$ in this plot.
}
\label{fig:HJ_J}
\end{center}
\end{figure}

\begin{figure}[t]
\begin{center}
\includegraphics[scale=1]{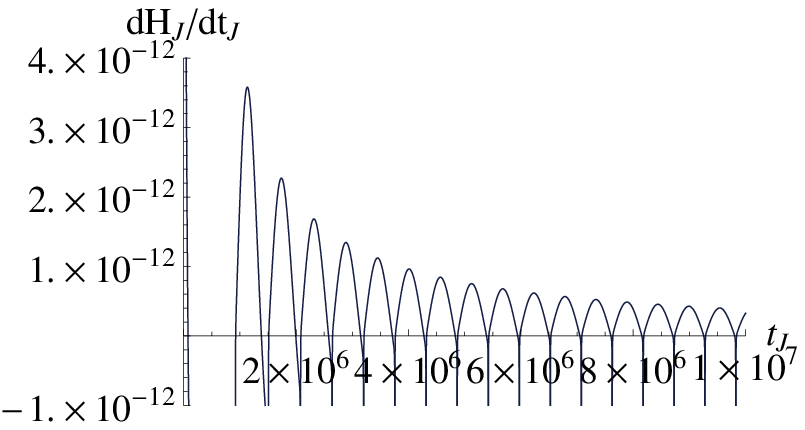}
\includegraphics[scale=1]{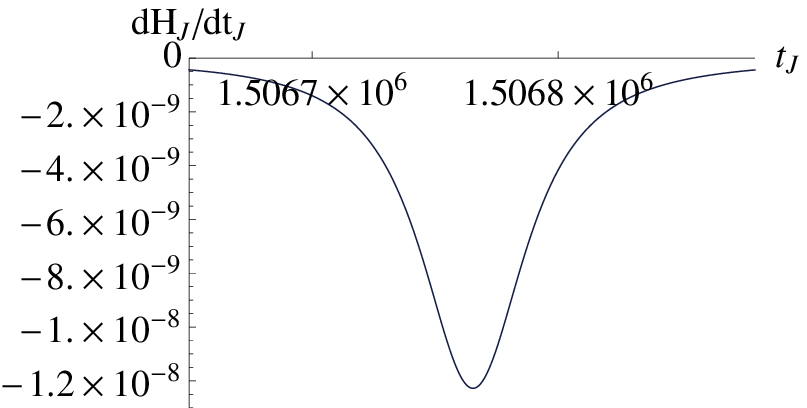}
\caption{ \small
Time evolution of $\dot{H}_J$ for positive region (left) and negative region (right) at the beginning of Phase 1. 
Note the difference in the amplitude of the positive peaks ($\dot{H}_J \sim 10^{-12}$) and 
negative peaks ($\dot{H}_J \sim 10^{-8}$). 
The height of the positive peaks of $\dot{H}_J$ is proportional to $t_J^{-1}$.
We take $M_\text{P} = 1$ in this plot.
}
\label{fig:dHJ}
\end{center}
\end{figure}

The soltuion~\eqref{eq:app:infsol} is no longer valid after $\Phi_J$ 
drops down to $\Phi_J \sim M_\text{P}/\sqrt{\xi}$.
Then, the inflaton $\phi_J$ starts to oscillate around the minimum of the potential.
First we show the numerical solutions in this phase in 
Figs.~\ref{fig:phiJ_dphiJ_tJ}--\ref{fig:dHJ}.
As we can see from them, the dynamics of $\phi_J$ and $H_J$ is peculiar in the Jordan frame.
In particular, $\dot{\phi}_J$ blows up at around the origin, and $H_J$ violently oscillates with time.
In the following, we see why these features appear in the Jordan frame.

First we study the dynamics of $\phi_J$.
In order to understand it qualitatively, 
we neglect the Hubble friction term in Eq.~\eqref{eq:phi_tEOM_2}, 
and hence Eq.~\eqref{eq:phi_tEOM_2} can be reduced as
\begin{align}
\tilde{\phi}''
+ \frac{\tilde{\phi}'^2 + \tilde{\phi}^2}{1 + \tilde{\phi}^2}\tilde{\phi}
&= 0,
\;\;\;\;\;\;
\tilde{\phi}
\equiv \frac{\sqrt{\xi(1 + 6\xi)}}{M_\text{P}} \phi_J,
\;\;\;\;\;\;
z
\equiv \frac{\lambda^{1/2}}{\sqrt{\xi(1 + 6\xi)}}M_\text{P}t_J,
\label{eq:app:phitildeEOM}
\end{align}
where the prime denotes the derivative with respect to $z$.
The amplitude satisfies $\tilde{\Phi} \gg 1$ in Phase~1.
We consider the cases with $\tilde{\phi}^2 \gg 1$ and $\tilde{\phi}^2 \ll 1$
separately in the following.
For $\tilde{\phi}^2 \gg 1$, we can approximate Eq.~\eqref{eq:app:phitildeEOM} as
\begin{align}
\tilde{\phi}''
+ \left( 1 + \frac{\tilde{\phi}'^2}{\tilde{\phi}^2} \right)\tilde\phi
&\simeq 0,
\end{align}
and the solution is
\begin{align}
\tilde{\phi}
&\simeq \pm \tilde{\Phi} \sqrt{\left| \cos (\sqrt{2} z) \right|},
\label{eq:app:phitilde_sol}
\end{align}
where the sign flips when $\tilde{\phi}$ crosses the origin.
Thus, the velocity is $\lvert\tilde{\phi}'\rvert \sim \tilde{\Phi}$
for $\tilde{\phi}^2 \gg 1$.
As $\tilde{\phi}$ approaches unity (or $\phi_J \sim M_\text{P}/\xi$), 
the above approximation breaks down.
At this time, the time variable satisfies 
$\lvert z - z_\times\rvert \sim \tilde{\Phi}^{-2}$, 
where $z_\times$ is the time when $\tilde{\phi}$ crosses the origin.
Then we find that the velocity is $\lvert \tilde{\phi}'\rvert \sim \tilde{\Phi}^2$
at $\tilde{\phi} \sim 1$. 
It means that the velocity blows up at around the origin since $\tilde{\Phi} \gg 1$ in Phase~1.
For $\tilde{\phi}^2 \ll 1$, we can approximate Eq.~\eqref{eq:app:phitildeEOM} as
\begin{align}
\tilde{\phi}''
+ \tilde{\phi}'^2 \tilde{\phi}
&= 0.
\end{align}
Note that $\tilde{\phi}'^2 \gg 1$ is already satisfied at $\tilde{\phi} \sim 1$ as we saw above.
It is easy to integrate the above equation to obtain
\begin{align}
\tilde{\phi}' 
&= Ce^{-\tilde{\phi}^2/2},
\end{align}
with $C$ being some integration constant.
It shows that $\tilde{\phi}'$ is roughly of the same order for $\tilde{\phi}^2 \ll 1$.
In summary, in terms of the original field $\phi_J$, the velocity shows the following property:
\begin{align}
	\dot{\phi}_J
&\sim
\left\{
\begin{matrix}
\;
\displaystyle\frac{\sqrt{\lambda}}{\xi}M_\text{P}\Phi_J
~~~~~~
\mathrm{for}
~~
\displaystyle \phi^2_J \gg M_\text{P}^2/\xi^2,
\\
\;\;
\displaystyle\sqrt{\lambda}\Phi_J^2
~~~~~~
\mathrm{for}
~~
\displaystyle \phi^2_J \ll M_\text{P}^2/\xi^2,
\end{matrix}
\right.
\label{eq:dphi_J}
\end{align}
and hence it blows up at around the origin in Phase~1.
It corresponds to the sharp spike-like features 
in the right panel of Fig.~\ref{fig:phiJ_dphiJ}.
The timescale of the whole oscillation $\Delta t_\mathrm{osc}$ 
is determined by the region $\phi_J^2 \gg M_\text{P}^2/\xi^2$:
\begin{align}
	m_\mathrm{osc} \equiv \Delta t_\mathrm{osc}^{-1} \sim \frac{\sqrt{\lambda}M_\text{P}}{\xi},
\end{align}
while the timescale $\Delta t_\mathrm{sp}$ for $\phi_J$ to pass the region $\phi_J^2 \lesssim M_\text{P}^2/\xi^2$ is given as
\begin{align}
	m_\mathrm{sp} \equiv \Delta t_\mathrm{sp}^{-1} \sim \frac{\sqrt{\lambda}\xi\Phi_J^2}{M_\text{P}}.
\end{align}
Here it is instructive to compare Eq.~\eqref{eq:dphi_J} with the dynamics in the Einstein frame.
Noting that $\Phi \sim \xi \Phi_J^2/M_\text{P}$ in Phase~1, we can see that 
the velocities in the Jordan frame and the Einstein frame satisfy
$\dot{\phi}_J \sim \sqrt{\lambda}M_\text{P} \Phi/\xi \sim \dot{\phi}$ for $\phi_J^2 \ll M_\text{P}^2/\xi^2$.
In other words, for $\phi_J^2 \gg M_\text{P}^2/\xi^2$, $\phi_J$ moves slower than that in the Einstein frame.
In order to catch up with $\dot{\phi}$ at around the origin, 
the mass term in the Jordan frame $m_{J\mathrm{eff}}$ must blow up at $\phi_J \lesssim M_\text{P}/\xi$.
This is one way to understand the spike-like behavior of $m_{J\mathrm{eff}}$ as we saw in the main text
(see Fig.~\ref{fig:phi_mJ2}).

Next we consider the dynamics of the Hubble parameter $H_J$.
In Fig.~\ref{fig:HJ_J}, we can see that 
$H_J$ violently oscillates with time (\textcolor{dark_blue}{blue line}).
Actually, it is often the case if we couple the inflaton to the gravity sector 
nonminimally~\cite{Jinno:2013fka,Ema:2015dka,Ema:2015oaa,Ema:2015eqa}.
In such a case, it is helpful to consider the following quantity~\cite{Ema:2015eqa}:\footnote{
It satisfies $\dot{J} \sim {\mathcal O}(H_JJ)$ even when
$\dot{H}_J \sim {\mathcal O}(m_{J\mathrm{eff}}H_J)$.
}
\begin{align}
J
&\equiv -\frac{1}{6M_\text{P}^2}{\mathcal L}_H
= H_J + \xi \frac{H_J\phi_J^2}{M_\text{P}^2} + \xi \frac{\phi_J\dot{\phi}_J}{M_\text{P}^2},
\label{eq:app:J}
\end{align}
where ${\mathcal L}_H$ is the derivative of the Lagrangian with respect to $H_J$.\footnote{
	We should perform the integration by parts to eliminate $\dot{H}_J$ in the Lagrangian
	before taking the derivative with respect to $H_J$.
}
In fact, in Fig.~\ref{fig:HJ_J}, we can see that $J$ has 
a suppressed amplitude of oscillation (\textcolor{dark_red}{red line}).
Thus, the oscillating mode in $H_J$ is estimated as\footnote{
	The second term in the R.H.S. of Eq.~(\ref{eq:app:J}) 
	is always smaller than the third term.
}
\begin{align}
\delta H_J
&\simeq -\xi \frac{\phi_J \dot{\phi}_J}{M_\text{P}^2}.
\end{align}
It always satisfies $\delta H_J \sim H_J$ irrespective of 
whether $\phi_J \lesssim M_\text{P}/\xi$ or $\phi_J \gtrsim M_\text{P}/\xi$.
It is consistent with the behavior of $H_J$ in Fig.~\ref{fig:HJ_J}.
The mass scales $m_\mathrm{sp}$ and $m_\mathrm{osc}$ determine the behavior of $\dot{H_J}$:
\begin{align}
	\dot{H}_J
&\sim
\left\{
\begin{matrix}
\;
\displaystyle m_\mathrm{osc}H_J
\sim
\frac{\lambda \xi \Phi_J^4}{M_\text{P}^2}
~~~~~~
\mathrm{for}
~~
\displaystyle \phi^2_J \gg M_\text{P}^2/\xi^2,
\vspace{1mm} \\
\;
\displaystyle m_\mathrm{sp}H_J
\sim
\frac{\lambda \Phi_J^2}{\xi}
~~~~~~
\mathrm{for}
~~
\displaystyle \phi^2_J \ll M_\text{P}^2/\xi^2,
\end{matrix}
\right.
\end{align}
which is consistent with Fig.~\ref{fig:dHJ}.

%%%%%%%%%%%%%%%%
%%%%%%%%%%%%%%%%

%%%%%%%%%%%%%%%%
%%%%%%%%%%%%%%%%

%%%%%%%%%%%%%%%%
%%%%%%%%%%%%%%%%

%%%%%%%%%%%%%%%%%%%%%%%%%%%%%%%%%%%%%%%%%%%%%%%%%%
\subsubsection*{Phase 2: $\Phi \ll M_\text{P} / \xi$}
%%%%%%%%%%%%%%%%%%%%%%%%%%%%%%%%%%%%%%%%%%%%%%%%%%

After the amplitude $\Phi_J$ drops below $M_\text{P}/\xi$,
the nonminimal term becomes ineffective.
Then, the background dynamics reduces 
to that in the case of the minimal Einstein gravity.

%%%%%%%%%%%%%%%%%%%%%%%%%%%%%%%%%%%%%%%%%%%%%%%%%%
\section{Particle production by spike-like external force}
\label{app:pp}
\setcounter{equation}{0}
%%%%%%%%%%%%%%%%%%%%%%%%%%%%%%%%%%%%%%%%%%%%%%%%%%

%%%%%%%%%%%%%%%%%%%%%%%%%%%%%%%%%%%%%%%%%%%%%%%%%%
\subsection{Basics of particle production}
%%%%%%%%%%%%%%%%%%%%%%%%%%%%%%%%%%%%%%%%%%%%%%%%%%

Here we summarize some basic formulae of particle production with a time-dependent mass term.
We consider the following equation of motion
\begin{align}
\ddot{\chi}(t,\bm{k})
+ \omega_k^2(t) \chi(t,\bm{k})
&= 0,
~~~
\omega_k^2(t) = k^2 + m_\chi^2(t),
\end{align}
where $\chi(t, \bm{k})$ is the Fourier component of 
a real canonical scalar field $\chi(t, \bm{x})$:
\begin{align}
\chi(t,\bm{x})
&= \int \frac{\dd^3k}{(2\pi)^3} e^{i \bm{k}\cdot \bm{x}} \chi(t,\bm{k}).
\end{align}
We expand $\chi(t, \bm{k})$ by the creation/annihilation operators as
\begin{align}
\chi(t,\bm{k})
&= \chi_k(t)a_{\bm{k}} + \chi^*_k(t)a^\dagger_{-{\bm k}},
\end{align}
where $a_{\bm{k}}$ and $a_{\bm{k}}^\dagger$ satisfies
\begin{align}
[a_{\bm k}(t),a^\dagger_{{\bm k}'}(t)]
&= (2\pi)^3\delta^3({\bm k} - {\bm k}'),
\;\;\;\;
[a_{\bm k}(t),a_{{\bm k}'}(t)]
= 
[a^\dagger_{\bm k}(t),a^\dagger_{{\bm k}'}(t)]
= 0.
\label{eq:app:commutation}
\end{align}
In order to be consistent with the canonical commutation relation of $\chi$,
\begin{align}
[\chi(t,{\bm k}), \dot{\chi}(t,-{\bm k'})]
= i(2\pi)^3\delta^3({\bm k} - {\bm k}'),
\end{align}
the mode function must satisfy the following Wronskian condition:
\begin{align}
\chi_k(t)\dot{\chi}_k^*(t)
- 
\chi_k^*(t)\dot{\chi}_k(t)
&= i.
\label{eq:Wronskian}
\end{align}
Now we express the mode function as
\begin{align}
\chi_k(t)
=
\frac{1}{\sqrt{2\omega_k}}
\left[
\alpha_k(t)e^{-i \int_0^t \dd t' \omega_k(t')}
+ 
\beta_k(t)e^{i \int_0^t \dd t' \omega_k(t')}
\right],
\end{align}
where $\alpha_k(t)$ and $\beta_k(t)$ are the so-called Bogoliubov coefficients.
There is a functional degree of freedom for the choice of $\alpha_k(t)$ and $\beta_k(t)$,
and we take it such that the coefficients satisfy
\begin{align}
\dot{\alpha}_k(t)
&= \frac{\dot{\omega}_k}{2\omega_k}e^{2i\int_0^t \dd t'\omega_k(t')}\beta_k(t),
\;\;\;\;
\dot{\beta}_k(t)
= \frac{\dot{\omega}_k}{2\omega_k}e^{-2i\int_0^t \dd t'\omega_k(t')}\alpha_k(t).
\label{eq:app:EOM_alphabeta}
\end{align}
The initial conditions are $\alpha_k(0) = 1$ and $\beta_k(0) = 0$ 
since the vacuum initial condition annihilated by $a_{\bm{k}}$ is given by
\begin{align}
\chi_k(t \to 0)
&\simeq \frac{1}{\sqrt{2\omega_k}}e^{-i\int_0^t \dd t' \omega_k(t')},
\;\;\;\;
\dot{\chi}_k(t \to 0)
\simeq -i \sqrt{\frac{\omega_k}{2}}e^{-i\int_0^t \dd t' \omega_k(t')}.
\end{align}
The Wronskian condition reads
\begin{align}
	\lvert \alpha_k(t) \rvert^2 - \lvert \beta_k(t) \rvert^2 = 1.
\end{align}
The number density and the energy density of $\chi$ are given by
\begin{align}
n_\chi(t)
= \int \frac{\dd^3k}{(2\pi)^3}f_\chi(t,k),
~~~
\rho_\chi(t)
= \int \frac{\dd^3k}{(2\pi)^3} \omega_k f_\chi(t,k),
\end{align}
where the occupation number $f_\chi(t,k)$ is defined as
\begin{align}
f_\chi(t,k)
&\equiv \frac{1}{2\omega_k}(|\dot{\chi}_k(t)|^2 + \omega_k^2|\chi_k(t)|^2)
- \frac{1}{2}
= |\beta_k(t)|^2.
\end{align}
Thus, we only have to integrate Eq.~\eqref{eq:app:EOM_alphabeta} to obtain the produced amount of $\chi$.
It is common in literature to use $\alpha_k(t)$ and $\beta_k(t)$ to study particle production.
However, in our case, it is of some help to define
\begin{align}
A_k(t)
&\equiv 
\alpha_k(t) e^{-i\int_0^t \dd t'\omega_k(t')},
\;\;\;\;
B_k(t)
\equiv 
\beta_k(t) e^{i\int_0^t \dd t'\omega_k(t')}.
\end{align}
In terms of them, the time evolution is described as
\begin{align}
\dot{A}_k(t)
&= \frac{\dot{\omega}_k}{2\omega_k}B_k(t) - i \omega_k(t) A_k(t),
\;\;\;\;
\dot{B}_k(t)
= \frac{\dot{\omega}_k}{2\omega_k}A_k(t) + i \omega_k(t) B_k(t).
\label{eq:app:EOM_AB}
\end{align}
The occupation number is also expressed as
\begin{align}
	f_\chi(t,k) = \lvert B_k(t)\rvert^2.
\end{align}
%%

%%%%%%%%%%%%%%%%%%%%%%%%%%%%%%%%%%%%%%%%%%%%%%%%%%
\subsection{Particle production by spike-like external force}
%%%%%%%%%%%%%%%%%%%%%%%%%%%%%%%%%%%%%%%%%%%%%%%%%%

Now let us consider particle production by spike-like external force.
We assume the time-varying mass to have the following form
\begin{align}
&m_\chi^2(t)
= \bar{m}_\chi^2 \; {\rm sp}(m_{\rm sp}t).
\label{eq:app:spike}
\end{align}
Here ${\rm sp}(m_{\rm sp}t)$ denotes a ``spike function'' whose maximal value is unity at $t = 0$,\footnote{
	We took $t = 0$ as the initial time in the previous subsection,
	but it corresponds to $t = -\infty$ here.
}
with $m_{\rm sp}$ being the inverse timescale of the spike.
We assume $\bar{m}_\chi \lesssim m_{\rm sp}$,\footnote{
In the model we analyze in the main text, the spike height is comparable to
its timescale, $\bar{m}_\chi \sim m_{\rm sp}$.
This makes the numerical results presented there somewhat deviate from the analytic estimation in this Appendix.
}
and ${\rm sp}(m_\mathrm{sp}t) \to 0$ for $\lvert m_\mathrm{sp}t\rvert \to \infty$.
Below we consider the modes with $k \ll \bar{m}_\chi$ and $k \gg \bar{m}_\chi$ separately.
This is because we have to take different strategy to study these cases analytically.
In particular, the Born approximation~\eqref{eq:app:beta_integrate_approx} 
is valid only for $k \gg \bar{m}_\chi$.
The result is summarized in Fig.~\ref{fig:occ},
and essentially the same result is also derived in App.~C of Ref.~\cite{Amin:2015ftc}.

%%%%%%%%%%%%%%%%%%%%%%%%%%%%%%%%%%%%%%%%%%%%%%%%%%
\subsubsection*{Case $k \ll \bar{m}_\chi$}
%%%%%%%%%%%%%%%%%%%%%%%%%%%%%%%%%%%%%%%%%%%%%%%%%%

\begin{figure}[t]
\begin{center}
\includegraphics[scale=0.85]{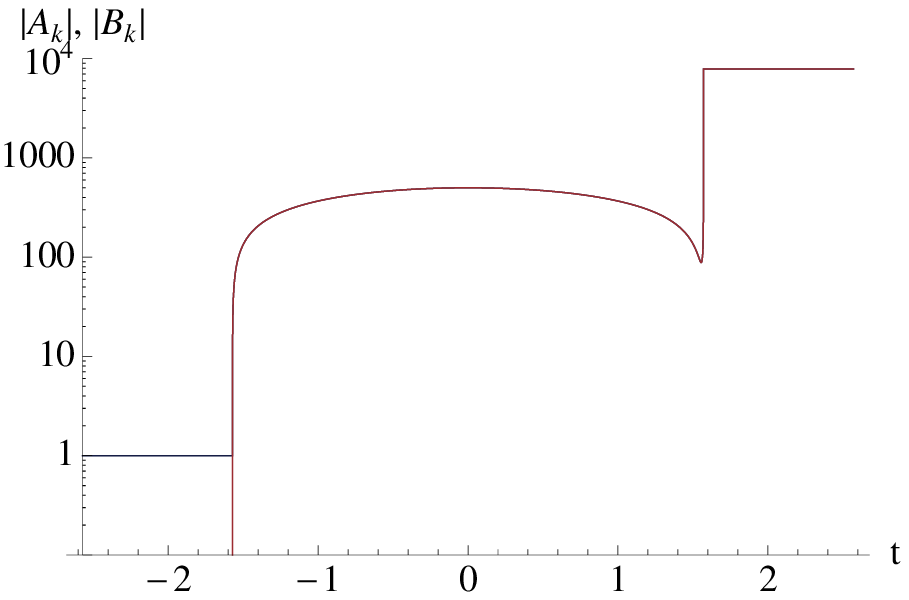}
\end{center}
\begin{center}
\includegraphics[scale=0.85]{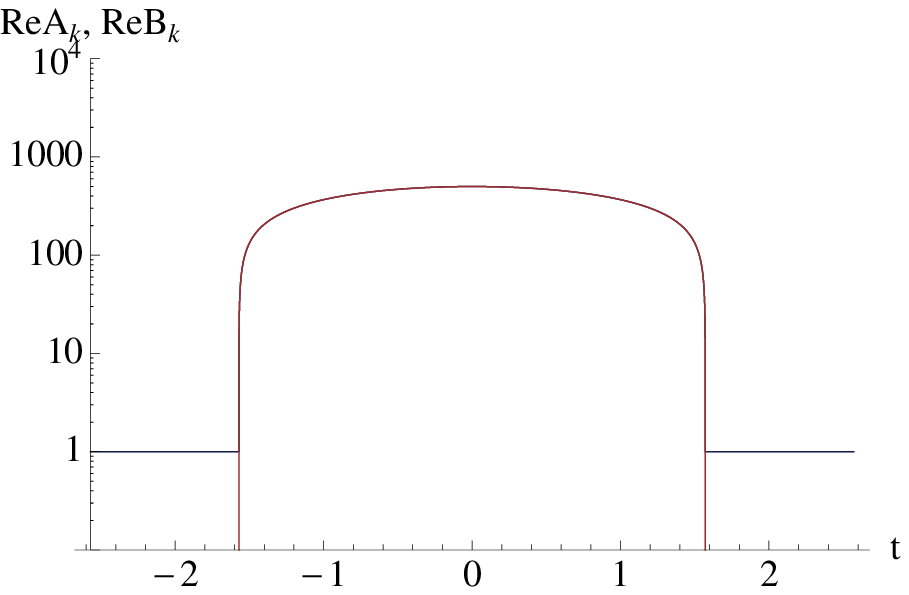}
\includegraphics[scale=0.85]{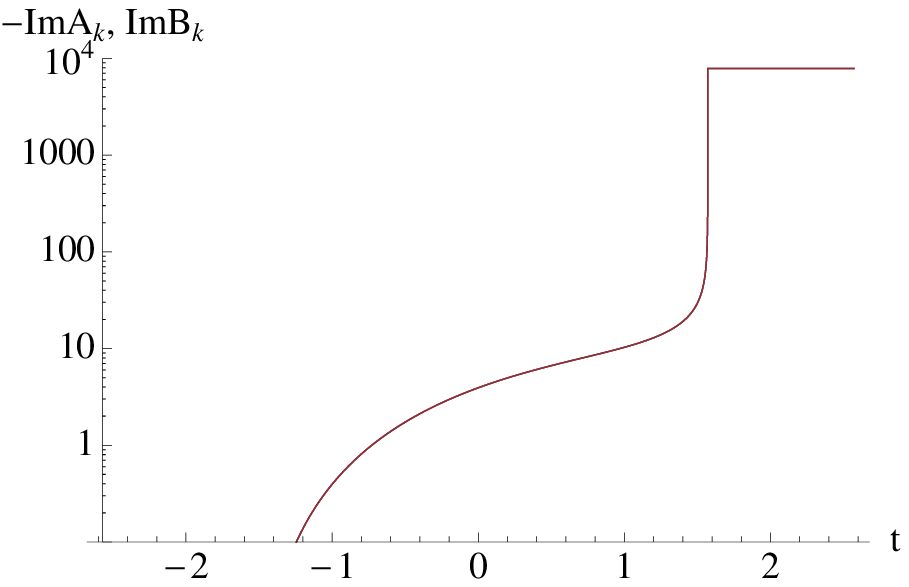}
\caption{ \small
Time evolution of $A_k$ (\textcolor{dark_blue}{blue}) and $B_k$ (\textcolor{dark_red}{red})
for $k = 10^{-8}$ with a trigonometric spike with $m_{\rm sp} = 1$ and 
$\bar{m}_\chi = 10^{-2}$  (see Eq.~(\ref{eq:sp_simple})).
Panels below show the real and imaginary parts, respectively,
while the above panel show the absolute value.
}
\label{fig:AB}
\end{center}
\end{figure}

First we consider low momentum modes with $k \ll \bar{m}_\chi$.
In this case, it is useful to consider $A_k$ and $B_k$ instead of $\alpha_k$ and $\beta_k$.
First note that the second terms in the R.H.S. of Eq.~(\ref{eq:app:EOM_AB}) cause
only ${\mathcal O}(\bar{m}_\chi / m_{\rm sp}) \ll 1$ phase rotation 
for $\lvert m_\mathrm{sp}t\rvert \lesssim 1$.
If we neglect these terms, it has the solution
\begin{align}
A_k(t)
&\simeq 
\cosh
\left[ 
\frac{1}{2} \ln \left( \frac{\omega_k}{k} \right)
\right],
\;\;\;\;
B_k(t)
\simeq 
\sinh
\left[ 
\frac{1}{2} \ln \left( \frac{\omega_k}{k} \right)
\right].
\label{eq:app:AB_approx}
\end{align}
They grows to $\mathcal{O}(\sqrt{\bar{m}_\chi/k}) \gg 1$ at around 
when $\mathrm{sp}(m_\mathrm{sp}t)$ becomes unity.
However, $B_k(t) \rightarrow 0$ as $m_\mathrm{sp}t \rightarrow \infty$,
and hence no net particle production occurs in Eq.~(\ref{eq:app:AB_approx}).
Thus, it is crucial to take the second terms in Eq.~(\ref{eq:app:EOM_AB}) into account.
Now we define ${\mathcal A}_k$ and ${\mathcal B}_k$ by
\begin{align}
A_k(t)
&\equiv 
\cosh
\left[ 
\frac{1}{2} \ln \left( \frac{\omega_k}{k} \right)
\right] + {\mathcal A}_k(t),
\;\;\;\;
B_k(t)
\equiv
\sinh
\left[ 
\frac{1}{2} \ln \left( \frac{\omega_k}{k} \right)
\right] + {\mathcal B}_k(t).
\end{align}
The evolution equations read
\begin{align}
\dot{{\mathcal A}}_k(t)
&= 
\frac{\dot{\omega}_k}{2\omega_k}{\mathcal B}_k(t)
- i \omega_k 
\cosh
\left[ 
\frac{1}{2} \ln \left( \frac{\omega_k}{k} \right)
\right] 
- i \omega_k 
{\mathcal A}_k(t),
\label{eq:app:mcAdot}
\\
\dot{{\mathcal B}}_k(t)
&= 
\frac{\dot{\omega}_k}{2\omega_k}{\mathcal A}_k(t)
+ i \omega_k 
\sinh
\left[ 
\frac{1}{2} \ln \left( \frac{\omega_k}{k} \right)
\right] 
+ i \omega_k 
{\mathcal B}_k(t).
\label{eq:app:mcBdot}
\end{align}
The initial conditions are
${\mathcal A}_k\rvert_{m_\mathrm{sp}t \rightarrow -\infty} 
= {\mathcal B}_k \rvert_{m_\mathrm{sp}t \rightarrow -\infty} = 0$.
In terms of them, we may see that the time evolution becomes as follows.
For $t \lesssim 0$, the second terms in the R.H.S. of 
Eqs.~(\ref{eq:app:mcAdot}) and (\ref{eq:app:mcBdot}) 
act as a source term for the imaginary parts of ${\mathcal A}_k$ and ${\mathcal B}_k$:
\begin{align}
{\rm Im}{\mathcal A}_k (t \sim 0),~
{\rm Im}{\mathcal B}_k (t \sim 0)
&\sim
\mp \int \dd t \; \omega_k \sqrt{\frac{\bar{m}_\chi}{k}} 
\sim 
\mp \frac{\bar{m}_\chi}{m_{\rm sp}} \sqrt{\frac{\bar{m}_\chi}{k}}.
\label{eq:mcAB_imaginary_0}
\end{align}
At around the spike function converges to zero, the first terms in the
R.H.S. of Eqs.~(\ref{eq:app:mcAdot}) and (\ref{eq:app:mcBdot}) 
give significant contribution.
Noting that $\dot{\omega}_k / \omega_k < 0$ and $-\mathcal{A}_k \sim \mathcal{B}_k$,
we may see that the first terms trigger significant growth
\begin{align}
{\rm Im} {\mathcal A}_k (t \gg m_{\rm sp}^{-1}),~
{\rm Im} {\mathcal B}_k (t \gg m_{\rm sp}^{-1})
&\sim 
\mp e^{\frac{1}{2} \left[ \ln (\omega_k/k) \right]_{t \gg m_{\rm sp}^{-1}}^{t \sim 0}}
\frac{\bar{m}_\chi}{m_{\rm sp}} \sqrt{\frac{\bar{m}_\chi}{k}}
\sim 
\mp  
\frac{\bar{m}_\chi^2}{m_{\rm sp}k}.
\label{eq:mcAB_imaginary_late}
\end{align}
In summary, the amount of particle production for $k \ll \bar{m}_\chi$ is estimated as
\begin{align}
f_\chi
&\sim |B_k|^2
\sim |{\mathcal B}_k|^2
\sim \frac{\bar{m}_\chi^4}{m_{\rm sp}^2k^2}.
\label{eq:app:f_low}
\end{align}
In Fig.~\ref{fig:AB} we show the time evolution of $A_k$ and $B_k$
for $k = 10^{-8}$ with a trigonometric spike with $m_{\rm sp} = 1$ and 
$\bar{m}_\chi = 10^{-2}$  (see Eq.~(\ref{eq:sp_simple})).
It is seen that the real parts behave as Eq.~(\ref{eq:app:AB_approx}),
while the imaginary parts show the time evolution explained in 
Eqs.~(\ref{eq:mcAB_imaginary_0})--(\ref{eq:mcAB_imaginary_late}).

%%%%%%%%%%%%%%%%
%%%%%%%%%%%%%%%%

%%%%%%%%%%%%%%%%%%%%%%%%%%%%%%%%%%%%%%%%%%%%%%%%%%
\subsubsection*{Case $k \gg \bar{m}_\chi$}
%%%%%%%%%%%%%%%%%%%%%%%%%%%%%%%%%%%%%%%%%%%%%%%%%%

\begin{figure}[t]
\begin{center}
\includegraphics[scale=0.9]{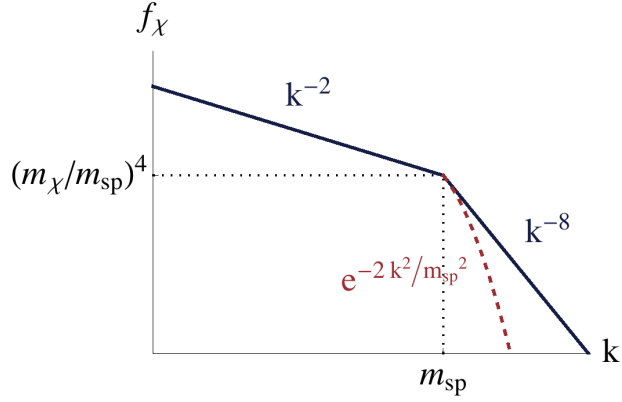}
\caption{ \small
Occupation number of $\chi$ produced by the spike-like external force~(\ref{eq:app:spike}).
}
\label{fig:occ}
\end{center}
\end{figure}

In this case, we can rely on the Born approximation.
By assuming $\alpha_k \simeq 1$, we can easily integrate Eq.~\eqref{eq:app:EOM_alphabeta} to obtain
\begin{align}
\beta_k(t)
&= \int_{-\infty}^t \dd t'\;
\frac{\frac{\dd}{\dd t'}\omega^2_k(t')}{4\omega_k^2(t')}
e^{-2i\int_0^{t'} \dd t'' \; \omega_k(t'')} 
\simeq 
\frac{i}{2k}
\int_{-\infty}^t \dd t'\;
m_\chi^2 (t')
e^{-2ikt'}.
\label{eq:app:beta_integrate_approx}
\end{align}
It is still difficult to analytically solve Eq.~\eqref{eq:app:beta_integrate_approx} in realistic cases
since $\mathrm{sp}(m_\mathrm{sp}t)$ is expressed by $\phi_J$ in a complicated manner.
Thus, here we assume a simpler form for $\mathrm{sp}(m_\mathrm{sp}t)$ to capture
generic features of particle production caused by the spike-like external force.
We consider two examples: (1) trigonometric function and (2) gaussian function
\begin{align}
(1):~ {\rm sp}(t)
&=
\left\{
\begin{matrix}
\cos^2(m_{\rm sp} t)
&~~
\mathrm{for}
~
\lvert m_{\rm sp}t \rvert < \pi/2 \vspace{1mm}, \\
0
&~~
\mathrm{otherwise},
\end{matrix}
\right.
~~~~
(2):~
{\rm sp}(t)
=
e^{-(m_\mathrm{sp}t)^2}.
\label{eq:sp_simple}
\end{align}
In these cases, it is easy to integrate Eq.~(\ref{eq:app:beta_integrate_approx}) to obtain
\begin{align}
(1):~ 
\beta_k
\simeq \frac{i\bar{m}_\chi^2}{4k^2} \frac{\sin(\pi k/m_{\rm sp})}{1 - k^2/m_{\rm sp}^2},
~~~~
(2):~
\beta_k
\simeq
\frac{i\sqrt{\pi}}{2}\frac{\bar{m}_\chi^2}{km_\mathrm{sp}}
e^{-k^2/m_\mathrm{sp}^2},
\label{eq:beta_trig_gauss}
\end{align}
and hence the occupation number is given by\footnote{
At $k \simeq m_{\rm sp}$, this expression of $f_\chi$ overestimates by a factor of ${\mathcal O}(0.1)$,
as one sees from Eq.~(\ref{eq:beta_trig_gauss}) with $k = m_{\rm sp}$.
}
\begin{align}
&f_\chi
\sim
|\beta_k|^2
\sim
\frac{\bar{m}_\chi^4}{m_{\rm sp}^2k^2}
~~~~
\mathrm{for}
~
k \ll m_\mathrm{sp},
\label{eq:app:f_high}
\end{align}
and a cut-off for $k \gg m_\mathrm{sp}$. The cut-offs are $f_\chi \propto k^{-8}$ for 
the trigonometric function and $f_\chi \propto e^{-k^2/m_\mathrm{sp}^2}$ for the gaussian function, respectively.
Note that the momentum dependence is the same for $k \ll \bar{m}_\chi$ and $\bar{m}_\chi \ll k \ll m_\mathrm{sp}$.
Fig.~\ref{fig:occ} shows the qualitative behavior of the occupation number.
Although we consider only two simple examples here, 
we may expect that Eq.~\eqref{eq:app:f_high} and a cut-off
behavior are generic features of the particle production by the spike-like external force.

%%%%%%%%%%%%%%%%
%%%%%%%%%%%%%%%%

%%%%%%%%%%%%%%%%%%%%%%%%%%%%%%%%%%%%%%%%%%%%%%%%%%
\section{Frame/gauge independence of particle production}
\label{app:frame_gauge}
\setcounter{equation}{0}
%%%%%%%%%%%%%%%%%%%%%%%%%%%%%%%%%%%%%%%%%%%%%%%%%%

%%%%%%%%%%%%%%%%%%%%%%%%%%%%%%%%%%%%%%%%%%%%%%%%%%
\subsection{Frame independence of particle production}
\label{app:frame}
\setcounter{equation}{0}
%%%%%%%%%%%%%%%%%%%%%%%%%%%%%%%%%%%%%%%%%%%%%%%%%%

Here we show that the action for the light degrees of freedom in the Jordan frame
is equivalent to that in the Einstein frame we have studied in Sec.~\ref{sec_pp}.
Then, the produced amount of these degrees of freedom must be the same in both frames.
In this subsection we keep the subscript $E$ to represent quantities in the Einstein frame.

%%%%%%%%%%%%%%%%%%%%%%%%%%%%%%%%%%%%%%%%%%%%%%%%%%
\subsubsection*{General argument}
%%%%%%%%%%%%%%%%%%%%%%%%%%%%%%%%%%%%%%%%%%%%%%%%%%

Let us start with the Jordan frame action for a scalar field $\chi$,
\begin{align}
	S = \int \dd^4x \sqrt{-g_J}\left[ \frac{1}{2}M_\text{P}^2 \Omega^2(t) R_J -\frac{F^2(t)}{2}g^{\mu\nu}_J\partial_\mu\chi \partial_\nu\chi - \frac{1}{2} 
	m_\chi^2(t) \chi^2 \right],
	\label{S_jordan}
\end{align}
where $\Omega$, $F$ and $m_\chi$ are assumed to be time dependent.
Using the conformal time $\dd\tau$ and defining $\tilde\chi_J\equiv F(t)a_J(t) \chi$, this reduces to
\begin{align}
	S = \int \dd^3x \dd\tau \left[ \frac{a_J^4}{2}M_\text{P}^2 \Omega^2(t) R_J +\frac{1}{2}\left({\tilde\chi_J'}\right)^2 - \frac{1}{2}\left(\partial_i\tilde\chi_J\right)^2 
	- \frac{1}{2} \widetilde m_{\chi_J}^2(t) \chi_J^2 \right],
\end{align}
where
\begin{align}
	\widetilde m_{\chi_J}^2(\tau) = \frac{a^2_Jm_\chi^2}{F^2} - \frac{(Fa_J)''}{Fa_J}.  \label{mchi_J}
\end{align}
Note that the time dependent kinetic term $F(t)$ leads to the effective mass term after canonical normalization.

On the other hand, we can transform the Jordan frame action (\ref{S_jordan}) to the Einstein frame by the Weyl transformation (\ref{eq:B_conformal}).
Then we find the Einstein frame action for $\chi$, after defining $\tilde\chi_E\equiv \left(F(t)a_E(t)/\Omega(t)\right) \chi$, as
\begin{align}
	S = \int \dd^3x\dd\tau \left[ \frac{a_E^4}{2}M_\text{P}^2R_E +\frac{1}{2}\left({\tilde\chi_E'}\right)^2 - \frac{1}{2}\left(\partial_i\tilde\chi_E\right)^2 
	- \frac{1}{2} \widetilde m_{\chi_E}^2(t) \chi_E^2 \right],
\end{align}
where
\begin{align}
	\widetilde m_{\chi_E}^2(\tau) = \frac{a^2_E m_\chi^2}{F^2\Omega^2} - \frac{(Fa_E/\Omega)''}{Fa_E/\Omega}.  \label{mchi_E}
\end{align}
Since $\tilde \chi_J$ and $\tilde\chi_E$ are canonical in each frame, the production rate of $\tilde\chi$ is determined 
solely by the time dependence of the effective mass (\ref{mchi_J}) and (\ref{mchi_E}).
Noting that $a_E = \Omega a_J$, we easily find $\widetilde m_{\chi_J}(\tau) = \widetilde m_{\chi_E}(\tau)$.
Hence, as expected, the production rate is the same in both frames for any time dependent function $\Omega(t)$, $F(t)$ and $m_\chi(t)$.

However, interpretation may be a bit different between the two frames.
For example, for a minimal scalar $F(t)=1$ and constant $m_\chi(t)$, a (violent) particle production is caused by
a peculiar time dependence of the scale factor $a_J$ in the Jordan frame,
while it is due to a rapid change of the conformal factor $\Omega$ in the Einstein frame.

For the gauge boson production, in order to show the equivalence between Jordan and Einstein frame,
it is sufficient to notice that the gauge boson mass (\ref{mA_Ein}) in the Einstein frame
is the same as that in the Jordan frame since $a_E = \Omega a_J$.

Note also that in the action (\ref{S_jordan}) we assumed that the mixing between $\chi$ and the metric is small enough.
For example, we can consider the case of nonminimal coupling $\Omega^2 \supset \xi \chi^2$
which potentially leads to the kinetic mixing of $\chi$ and scalar part of the metric.
As long as $\sqrt{\xi^2 \left<\chi^2\right>} \ll M_\text{P} \lvert F\rvert$, such a mixing effect can be safely neglected.
If $\chi$ is the inflaton itself, the mixing is not neglected since $\xi \Phi \gg M_\text{P}$ after inflation.
Below we see that the inflaton self-production rate in the Jordan frame is the same as that in the Einstein frame
after taking the mixing into account.

%%%%%%%%%%%%%%%%%%%%%%%%%%%%%%%%%%%%%%%%%%%%%%%%%%
\subsubsection*{Self production}
%%%%%%%%%%%%%%%%%%%%%%%%%%%%%%%%%%%%%%%%%%%%%%%%%%

To show the equivalence of the inflaton self production rate between two frames may be a bit non-trivial
due to the mixing of inflaton fluctuation and the metric perturbation.
Here let us assume the action (\ref{eq:B_SJ}) and see the production of the scalar perturbation in the Jordan frame.
Again using the ADM formalism
\begin{align}
\dd s^2
&= -N_J^2\dd t^2 + a^2_Je^{2\zeta} (\dd x^i + \beta^i \dd t)(\dd x^i + \beta^i \dd t),
\end{align}
and taking the unitary gauge $\phi_J(t, \vec{x}) = \bar{\phi}_J(t)$,
we have~\cite{Kobayashi:2011nu}
\begin{align}
S
&= \int \dd^3x\dd\tau \; 
\frac{\bar{\phi}_J'^2 \Omega^2}{2a_J^2J^2}
\left[ 1 + \xi(1 + 6\xi)\frac{\bar{\phi}_J^2}{M_\text{P}^2} \right]
\left[
\zeta'^2 - (\nabla \zeta)^2 
\right],
\label{eq:app:zeta_J}
\end{align}
where the prime denotes the derivative with respect to the conformal time $\tau$,
the conformal factor $\Omega$ is given as Eq.~\eqref{eq:B_Omega},
and $J$ is the adiabatic invariant defined in Eq.~(\ref{eq:app:J}).
Note that $J$ is related to the Hubble parameter in the Einstein frame as
\begin{align}
J
&= \Omega^3 H_E,
\end{align}
where we have used $H_E = (1/a_E)(\dd a_E/\dd t_E)$, $\dd t_E = \Omega \dd t_J$ and $a_E = \Omega a_J$.
Then, we can rewrite the action~\eqref{eq:app:zeta_J} as
\begin{align}
S
&= \int \dd^3x\dd\tau \; 
\frac{\bar{\phi}_J'^2 }{2a_E^2H_E^2\Omega^2}
\left[ 1 + \xi(1 + 6\xi)\frac{\bar{\phi}_J^2}{M_\text{P}^2} \right]
\left[
\zeta'^2 - (\nabla \zeta)^2
\right],
\end{align}
Since $\bar{\phi}$ and $\bar{\phi}_E$ are related with each other through Eq.~(\ref{eq:B_dphi_tEdphi}),
the above action reduces to Eq.~(\ref{eq:pp_zeta_E_conformal}).
Thus, the production of $\zeta$ is the same 
both in the Einstein and Jordan frames.

%%%%%%%%%%%%%%%%%%%%%%%%%%%%%%%%%%%%%%%%%%%%%%%%%%
\subsection{Gauge independence of particle production}
\label{app:gauge}
%%%%%%%%%%%%%%%%%%%%%%%%%%%%%%%%%%%%%%%%%%%%%%%%%%

In this appendix, we show the gauge independence of particle production 
in the gauged U(1) inflaton case.
In particular, the unitary gauge used in the main text is ill-defined at the origin $\phi = 0$,
and hence we work in the Coulomb gauge which is well defined at the origin here.
We will confirm that the singularity of the unitary gauge at $\phi = 0$ does not spoil the discussion
in the main text by explicitly showing that the same result is obtained in the Coulomb gauge.

We consider the following action:
\begin{align}
S
&=
\int \dd^4x \sqrt{-g_J}
\left[ 
\left( \frac{M_\text{P}^2}{2} + \xi |\phi_J|^2 \right) R_J
- \frac{1}{4}g^{\mu\rho}_Jg^{\nu\sigma}_JF_{\mu \nu}F_{\rho \sigma}
- g^{\mu\nu}_J(D_\mu \phi_J)^\dagger (D_\nu \phi_J)
- V_J(|\phi_J|^2)
\right],
\label{eq:app:gauge_S}
\end{align}
where $F_{\mu\nu}$ is the field strength, $D_\mu = \partial_\mu - igA_\mu$, 
$A_\mu$ is the gauge field and $g$ is the gauge coupling, respectively.
We take $\phi_J = \left( \phi_{JR} + i\phi_{JI}\right)/\sqrt{2}$, and identify $\phi_{JR}$ as the inflaton.
In this case, $\phi_{JR}$ and the scalar component of the metric do not mix with $\phi_{JI}$ and the gauge boson.
Thus, we concentrate only on the quadratic action for $\phi_{JI}$ and the gauge bosons, and take $\phi_{JR} = \phi_{JR}(t)$.
We take the Coulomb gauge $\partial_i A_i = 0$, and then the quadratic action is written as
\begin{align}
	S_\mathrm{quad} = 
	\int \dd^3x \dd\tau
	&\left[
	-\frac{1}{4}\eta^{\mu\rho}\eta^{\nu\sigma} F_{\mu\nu}F_{\rho\sigma} 
	-\frac{m_A^2}{2}\eta^{\mu\nu} A_\mu A_\nu
	-\frac{a_J^2}{2}\eta^{\mu\nu}\partial_\mu \phi_{JI} \partial_\nu \phi_{JI} 
	\right. \nonumber \\
	&~~ -\left.
	\frac{a_J^4}{2}\left(\lambda \phi_{JR}^2 - \xi R_J\right)\phi_{JI}^2
	- a_J^2g A_\tau \left(\phi_{JI}\phi_{JR}' - \phi_{JR}\phi_{JI}'\right)
	\right],
\end{align}
where the prime denotes derivatives with respect to the conformal time $\tau$, 
and $m_A^2 \equiv g^2 a_J^2 \phi_{JR}^2$. Note that this definition of $m_A$ is 
equivalent to Eq.~\eqref{mA_Ein} since $a_J = a_E/\Omega$.
In the Coulomb gauge where the longitudinal mode is gauged away, 
the kinetic term for the gauge boson is given as
\begin{align}
	-\frac{1}{4}\eta^{\mu\rho}\eta^{\nu\sigma}F_{\mu\nu}F_{\rho\sigma}
	=
	\frac{1}{2}\left(\partial_i A_\tau\right)^2 + \frac{1}{2}\left\lvert \vec{A}_T' \right\rvert ^2
	- \frac{1}{2}\left\lvert \vec{\nabla} \times \vec{A}_T\right\rvert^2.
\end{align}
Then, in the momentum space, the quadratic action reads
\begin{align}
	S_\mathrm{quad} = S_{A_T} + S_\mathrm{G},
\end{align}
where
\begin{align}
	S_{A_T} 
	&= \frac{1}{2}\int \frac{\dd\tau \dd^3k}{\left(2\pi\right)^3}
	\left[\left\lvert\vec{A}_T'\right\rvert^2 
	- \left(k^2 + m_A^2\right)\left\lvert \vec{A}_T\right\rvert^2
	\right],
\end{align}
and
\begin{align}
	S_\mathrm{G}
	=
	\int \frac{\dd\tau \dd^3k}{\left(2\pi\right)^3}
	&\left[
	\frac{m_A^2 + k^2}{2}\left\lvert A_\tau - \frac{a_J^2 g}{m_A^2 + k^2}
	\left(\phi_{JI}\phi_{JR}' - \phi_{JR}\phi_{JI}'\right) \right\rvert^2
	+ \frac{a_J^2}{2}\left(\left\lvert \phi_{JI}'\right\rvert^2 - k^2 \left\lvert \phi_{JI}\right\rvert^2 \right)
	\right. \nonumber \\
	&- \left.
	\frac{a_J^4g^2}{2\left(m_A^2 + k^2\right)}
	\left\lvert \phi_{JI}\phi_{JR}' - \phi_{JR}\phi_{JI}'\right\rvert^2
	- \frac{a_J^4}{2}\left(\lambda \phi_{JR}^2 - \xi R_J\right) \left\lvert \phi_{JI}\right\rvert^2
	\right].
	\label{eq:action_goldstone}
\end{align}
Thus, the action for the transverse mode is the same as that in the unitary gauge.
From now we focus on the action for the Goldstone mode $S_\mathrm{G}$.
After integrating out $A_\tau$ and canonically normalizing 
$\phi_{JI}$ as $\phi_I \equiv a_J k \phi_{JI}/\sqrt{m_A^2 + k^2}$,
we rewrite the action~\eqref{eq:action_goldstone} as
\begin{align}
	S_G
	&= \frac{1}{2}\int \frac{\dd^3k \dd\tau}{\left(2\pi\right)^3}
	\left[
	\left\lvert \phi_I'\right\rvert^2 - \left(k^2 + m_A^2 + \delta m_{I}^2 \right)\left\lvert \phi_I \right\rvert^2
	\right],
	\label{eq:goldstone_gauge}
\end{align}
where
\begin{align}
	\delta m_{I}^2 
	=&
	\frac{a_J^2}{k^2}g^2 \left(\phi_{JR}'\right)^2 
	+ \frac{k^2 + m_A^2}{a_J^2 k^2}\frac{\dd}{\dd\tau}\left(\frac{a_J^4g^2\phi_{JR}\phi_{JR}'}{k^2 + m_A^2}\right) \nonumber \\
	&- \frac{\sqrt{k^2 + m_A^2}}{a_Jk} \frac{\dd^2}{\dd\tau^2}\left(\frac{a_J k}{\sqrt{k^2 + m_A^2}}\right)
	+ \frac{a_J^2 \left(k^2 + m_A^2\right)}{k^2}\left(\lambda \phi_{JR}^2 - \xi R_J\right).
\end{align}
By using the background equation of motion~\eqref{eq:eom1_J}, the last term is expressed as
\begin{align}
	\lambda \phi_{JR}^2 - \xi R_J 
	= 
	- \frac{1}{\phi_{JR}}\frac{\dd^2 \phi_{JR}}{\dd t_J^2}
	- \frac{3 H_J}{\phi_{JR}}\frac{\dd \phi_{JR}}{\dd t_J}.
\end{align}
Then, after some computation, we obtain
\begin{align}
	\delta m_I^2 = -\frac{k^2}{k^2 + m_A^2}\left(\frac{m_A''}{m_A} - \frac{3 \left(m_A'\right)^2}{k^2 + m_A^2}\right).
\end{align}
Thus, the action for the goldstone boson $\phi_I$ reduces to Eq.~\eqref{S_AL},
and hence we have proven that the action is the same between the unitary and Coulomb gauges.

%%%%%%%%%%%%%%%%%%%%%%%%%%%%%%%%%%%%%%%%%%%%%%%%%%
\small
\bibliography{ref}
%%%%%%%%%%%%%%%%%%%%%%%%%%%%%%%%%%%%%%%%%%%%%%%%%%

%%%%%%%%%%%%%%%%%%%%%%%%%%%%%%%%%%%%%%%%%%%%%%%%%%
\end{document}